\UseRawInputEncoding
\documentclass[final]{IEEEtran}

\usepackage{picinpar}
\usepackage{subfig}
\usepackage{subfloat}
\usepackage{setspace}
\usepackage[all]{xy}
\usepackage{color}
\DeclareCaptionFont{red}{\color{red}}
\DeclareCaptionFont{green}{\color{green}}
\DeclareCaptionFont{blue}{\color{blue}}
\usepackage{cite}
\usepackage[cmex10]{amsmath}
\usepackage{cuted}
\usepackage{graphicx}
\usepackage{picinpar}
\usepackage[cmex10]{amsmath}
\usepackage{amsmath,amsfonts,amssymb}
\usepackage{array}
\usepackage{mdwmath}
\usepackage{mdwtab}
\usepackage{eqparbox}
\usepackage{stfloats}
\usepackage{booktabs}
\usepackage{multirow}
\usepackage{etoolbox}
\usepackage{makecell}
\usepackage[justification=centering]{caption}
\usepackage{algorithmic}
\usepackage[ruled,lined,boxed]{algorithm2e}
\usepackage{fixltx2e}
\usepackage{bm}

\newcommand{\e}{\begin{equation}}
\newcommand{\ee}{\end{equation}}
\newcommand{\eqn}{\begin{eqnarray}}
\newcommand{\eeqn}{\end{eqnarray}}

\allowdisplaybreaks
\begin{document}
\title{Compressive Sensing Based Joint Activity and Data Detection for Grant-Free Massive IoT Access}

\author{Yikun Mei, Zhen Gao, Yongpeng Wu, Wei Chen, Jun Zhang, Derrick Wing Kwan Ng,~\IEEEmembership{Fellow,~IEEE}, Marco Di Renzo,~\IEEEmembership{Fellow,~IEEE}
\thanks{The work of Z. Gao was supported in part by the Natural Science Foundation of China (NSFC) under Grant 62071044; in part by the Beijing Natural Science Foundation under Grant L182024; and in part by the Basic Science Center for Autonomous Intelligent Unmanned Systems under Grant 62088101, and in part by the Science and Technology Innovation Plan from Beijing Institute of Technology. The work of Y. Wu was supported in part by the National Key R\&D Program of China under Grant 2018YFB1801102 and NSFC under Grant 62071289. The work of D. W. K. Ng was supported by funding from the UNSW Digital Grid Futures Institute, UNSW, Sydney, under a cross-disciplinary fund scheme and by the Australian Research Council's Discovery Project under Grant DP210102169. \emph{(Corresponding author: Zhen Gao.)}}
\thanks{Y. Mei and Z. Gao are with the School of Information and Electronics, Beijing Institute of Technology, Beijing 100081, China (e-mails: meiyikun@bit.edu.cn; gaozhen16@bit.edu.cn).}
\thanks{Y. Wu is with the Department of Electronic Engineering, Shanghai Jiao Tong University, Shanghai 200240, China (e-mail: yongpeng.wu@sjtu.edu.cn).}
\thanks{W. Chen is with the State Key Laboratory of Rail Traffic Control and Safety, Beijing Jiaotong University, Beijing 100044, China (e-mail: weich@bjtu.edu.cn).}
\thanks{D. W. K. Ng is with the School of Electrical Engineering and Telecommunications, University of New South Wales, Sydney, NSW 2025, Australia (e-mail: w.k.ng@unsw.edu.au). }
\thanks{M. D. Renzo is with the Laboratoire des Signaux et Syst\`emes, CNRS, CentraleSup\'elec, Univ Paris Sud, Universit\'e Paris-Saclay, 91192 Paris, France ¡ä(e-mail: marco.direnzo@centralesupelec.fr).}
}

\maketitle
\begin{abstract}
Massive machine-type communications (mMTC) are poised to provide ubiquitous connectivity for billions of Internet-of-Things (IoT) devices. However, the required low-latency massive access necessitates a paradigm shift in the design of random access schemes, which invokes a need of efficient joint activity and data detection (JADD) algorithms. By exploiting the feature of sporadic traffic in massive access, a beacon-aided slotted grant-free massive access solution is proposed. Specifically, we spread the uplink access signals in multiple subcarriers with pre-equalization processing and formulate the JADD as a multiple measurement vectors (MMV) compressive sensing problem. Moreover, to leverage the structured sparsity of uplink massive access signals among multiple time slots, we develop two computationally efficient detection algorithms, which are termed as orthogonal approximate message passing (OAMP)-MMV algorithm with simplified structure learning (SSL) and accurate structure learning (ASL). To achieve accurate detection, the expectation maximization algorithm is exploited for learning the sparsity ratio and the noise variance. To further improve the detection performance, channel coding is applied and successive interference cancellation (SIC)-based OAMP-MMV-SSL and OAMP-MMV-ASL algorithms are developed, where the likelihood ratio obtained in the soft-decision can be exploited for refining the activity identification. Finally, the state evolution of the proposed OAMP-MMV-SSL and OAMP-MMV-ASL algorithms is derived to predict the performance theoretically. Simulation results verify that the proposed solutions outperform various state-of-the-art baseline schemes, enabling low-latency random access and high-reliable massive IoT connectivity with overloading.
\end{abstract}

\begin{IEEEkeywords}
Compressive sensing, grant-free massive access, orthogonal approximate message passing, multiple measurement vectors, successive interference cancellation.
\end{IEEEkeywords}

\IEEEpeerreviewmaketitle

\section{Introduction}\label{S1}
Next-generation wireless communications promise the unprecedented connectivity for myriad Internet-of-Things (IoT) devices, which compose the prime driver of massive machine-type communications (mMTC) \cite{mMTC1,mMTC2}. In stark contrast to the conventional downlink-dominated human-type communications (HTC), the emerging mMTC are uplink-dominated having the features of sporadic traffic with low rates and short packets \cite{mMTC1,mMTC2,mMTC_feature,mMTC_new}. With this premise, mMTC compel a paradigm shift in the design of uplink random access procedures, which previously focused on HTC while were rarely prioritized for massive machine-type IoT devices \cite{mMTC_new}.

The current random access protocols, including long term evolution-advanced (LTE-A) and the fifth-generation new radio (5G NR) release 16, adopt the conventional grant-based access procedure \cite{3GPP}. In particular, grant-based random access usually requires a complicated handshaking protocol, through which a base station (BS) allocates orthogonal radio resources to different users for uplink transmission. However, such grant-based protocols suffer from low access efficiency in mMTC, since the payload size can be much smaller than the amount of the signaling overhead. Moreover, since the IoT devices can be massive, the contention for resources request will lead to inevitable collisions, which can dramatically increase the access latency \cite{mMTC_challenge1,mMTC_challenge2}. Therefore, an efficient massive random access paradigm is needed for enabling low-latency and high-reliability mMTC.

\vspace{-2mm}
\subsection{Related Work}\label{S1.1}

To accommodate the challenging requirements of mMTC, the grant-free massive access paradigm has been emerging as a promising solution and has drawn great attention \cite{{grant_free,MA5B}}. In grant-free random access schemes, users directly transmit their data to the BS without performing a sophisticated handshaking process. In this way, complicated control signaling procedures can be avoided, the access latency can be substantially reduced, and more radio resources can be saved for payload transmission. On the other hand, since the number of potential IoT devices in mMTC is generally massive, it is impossible to allocate orthogonal channel resources to all devices for facilitating data detection at the BS \cite{{NOMA1},{NOMA2}}. Fortunately, IoT devices, in most instances, remain inactive for energy saving and other reasons (e.g., data collection and infrequent alarm). In other words, only a small number of devices are active and need to access the BS at any given time.

In view of this fact, most grant-free access protocols adopt limited non-orthogonal radio resources to serve massively deployed IoT devices \cite{{JSAC_DY,CL_WBC1,CL_WBC2,TWC_DY,IORLS,TVT_SBL,CL_WC,TVT_Shim}}, where the detection of devices' activity and data is a challenging problem to be solved. Due to the feature of sporadic traffic in mMTC, the joint activity and data detection (JADD) procedure can be formulated as a sparse signal recovery problem, which can be handled by using compressive sensing (CS) algorithms. In fact, CS-based multi-user detectors (MUD) can date back to code division multiple access (CDMA) systems. In \cite{ZH}, for instance, MUDs based on the sparsity-exploiting maximum a posteriori probability (MAP) criterion were proposed for sparse CDMA systems. Besides, considering that the sparsity can be time-varying, CS-based nonlinear detectors and conventional linear detectors were combined for improving robustness \cite{CL_Shim}. Also, with the advent of mMTC, grant-free random access in conjunction with CS-based detectors were conceived for JADD in \cite{JSAC_DY,CL_WBC1,CL_WBC2,TWC_DY,IORLS,TVT_SBL,CL_WC,TVT_Shim}. In general, these CS-based detection algorithms can be divided into two categories: greedy-based algorithms and Bayesian inference-based algorithms \cite{{CS_OMP},{CS_SP},{AMP},{SBL},{EP}}. The former category was developed from the CS greedy algorithms including the orthogonal match pursuit (OMP) algorithm \cite{CS_OMP}, the subspace pursuit (SP) algorithm \cite{CS_SP}, etc., which heuristically find the most correlated atoms in each iteration. Furthermore, to exploit the temporal partial correlation of device activity, a modified OMP algorithm and a modified SP algorithm were proposed in \cite{JSAC_DY} and \cite{CL_WBC2}, respectively, where the estimated support was exploited as prior knowledge for the next detection. Moreover, considering that the temporal correlation of device activity typically remains unchanged during one frame, the structured iterative support detection algorithm and modified adaptive SP algorithm were proposed in \cite{CL_WBC1} and \cite{TWC_DY}, respectively, whereby the block sparsity of random access signals in multiple slots was leveraged for performance improvement. However, \cite{JSAC_DY,CL_WBC1,CL_WBC2,TWC_DY} fail to utilize the a priori distribution of the transmitted constellation symbols, and the calculation of matrix inversions that are needed in those algorithms pose an exceedingly high computational complexity when the number of devices becomes large.

Another category of CS-based random access detection algorithms are built on the framework of Bayesian inference, which includes the approximate messaging passing (AMP) algorithm \cite{AMP}, the sparse Bayesian learning (SBL) algorithm \cite{SBL}, the expectation propagation (EP) algorithm \cite{EP}, and etc. A common feature of them lies in the fact that the adopted estimators are designed based on the posterior distribution, which often depends on the a priori distribution of devices' symbols. Due to their flexibility and accuracy, they have become popular recently. In \cite{TVT_SBL}, for example, the modified SBL algorithms were applied for JADD, by making use of the sparse device activity and the prior knowledge of the signals' constellation. By resorting to the expectation-maximization (EM) algorithm, the knowledge of device activity is not required by the proposed algorithms. However, the associated computational complexity can be high since the SBL algorithm requires the computation of a matrix inversion in each iteration. By subtly using the likelihood ratio as the judgment metric for JADD, a MAP-based iterative detection scheme was proposed in \cite{TVT_Shim}. However, the noise variance and probability of devices' activity were assumed to be perfectly known at the BS, which can be impractical. To effectively reduce the computational complexity and speed up the convergence, an AMP-based detection scheme was proposed in \cite{CL_WC}, where the hyper-parameters of device activity can be learned through the EM algorithm. In \cite{CL_WC}, in addition, the sensing matrix consists of channel coefficients and spreading sequences. This indicates that the channel state information (CSI) is assumed to be available in advance at the BS. Moreover, the adopted spreading sequences require the elements to be independent
and identically distributed (i.i.d), which results in the design difficulty. To address those issues, our works aim at designing a practical JADD that guarantees high efficiency and accuracy.

\vspace{-2mm}
\subsection{Our Contribution}\label{S1.2}
In this paper, we propose a beacon-aided slotted grant-free massive access scheme for mMTC and investigate the JADD for application to orthogonal frequency division multiplexing (OFDM) systems\footnote{This article was presented in part at the International Conference on UK-China Emerging Technologies, Glasgow, UK, August 2020.}. Specifically, due to the sporadic traffic of machine-type IoT devices, the JADD problem at the BS can be formulated as a CS problem and can be solved by using the orthogonal AMP (OAMP) algorithm \cite{OAMP}. Moreover, the structured sparsity of uplink access signals in several successive OFDM symbols can be exploited by leveraging the multiple measurement vectors (MMV) CS algorithm for performance enhancement. However, the conventional OAMP algorithm is limited to the single measurement vector (SMV) CS problem and is incapable of solving the MMV CS problem with structured sparsity. To this end, we propose two detection algorithms that are referred to as OAMP-MMV algorithms with simplified structure learning (SSL) and accurate structure learning (ASL). In both of them, the prior knowledge of discrete constellation symbols is utilized, and EM algorithms are employed to estimate the sparsity ratio and noise variance for reliable detection performance. Even for the overloading case, the proposed OAMP-MMV algorithms can still achieve effective JADD. Meanwhile, the state evolution (SE) is derived for theoretically analyzing the performance of our proposed algorithms. Furthermore, channel coding and successive interference cancellation (SIC) are integrated into the proposed OAMP-MMV-SSL and OAMP-MMV-ASL algorithms for performance enhancement, where the likelihood ratio (LLR) is utilized for improving the activity identification. The main contributions made by this paper can be summarized as follows.

$\bullet$ \textbf{Beacon-aided slotted grant-free massive access scheme:}
Previous works, e.g., \cite{JSAC_DY,CL_WBC1,CL_WBC2,TWC_DY,IORLS,TVT_SBL,CL_WC,TVT_Shim}, mainly considered that the channels of all devices are known at the BS. In contrast, we remove this assumption and propose a beacon-aided random access scheme. After listening to the beacon that is periodically broadcast by the BS, the IoT devices perform synchronization, power control, and channel estimation. Then, a pre-equalization processing is introduced at the devices so that the uplink channels required by the BS are effectively avoided.

$\bullet$ \textbf{OAMP-MMV-SSL algorithm:}
Exploiting the structured sparsity in several successive time slots, we develop an OAMP-MMV-SSL algorithm, where the SSL strategy is proposed to improve the performance of activity detection. In particular, the discrete a priori distribution is assumed for characterizing the transmitted constellation symbols, and EM algorithm is incorporated to learn the unknown sparsity ratio and the noise variance. Besides, the SE is derived for characterizing the algorithm's performance.

$\bullet$ \textbf{OAMP-MMV-ASL algorithm:}
To fully exploit the aforementioned structured sparsity, we further propose an OAMP-MMV-ASL algorithm, where an ASL strategy with more sophisticated a priori distribution is considered. Particularly, by establishing a graphical model, the structured sparsity can be accurately described in the a priori distribution. We show that the OAMP-MMV-ASL algorithm can achieve better performance than OAMP-MMV-SSL algorithm at the cost of increasing the computational complexity.

$\bullet$ \textbf{SIC-based JADD:}
In contrast to most prior works, e.g., \cite{JSAC_DY,CL_WBC1,TWC_DY,IORLS,TVT_SBL,CL_WC,TVT_Shim}, which assume uncoded random access, our proposed access scheme takes channel coding into account and develop SIC-based OAMP-MMV-SSL and OAMP-MMV-ASL algorithms. Particularly, the LLR obtained in soft-decision is exploited for refining the active device identification.

Compared to the previous study \cite{CL_WC}, our proposed schemes are developed from the OAMP algorithm rather than the AMP algorithm, where the OAMP algorithm is more robust than the traditional AMP algorithm and has the less requirements on the sensing matrix. A similar point lies in that both [21] and our work employ the EM method to update the sparsity ratio. However, our proposed OAMP-MMV-ASL algorithm considers the more sophisticated a priori distributions than the method proposed in \cite{CL_WC} and resorts to sum-product rule for learning the sparsity structure, which has been verified by the numerical results to have the more accurate JADD performance. Besides, the noise variance is assumed to be known in \cite{CL_WC}, while we remove this impractical constraint and apply the EM algorithm to learn the unknown noise variance. Last but not least, the integration with channel coding and SIC is not considered in \cite{CL_WC}, and we will show that the proposed SIC-based OAMP-MMV algorithms effectively improve the performance compared with the OAMP-MMV algorithms without SIC processing.

\color{black}
The rest of the paper is organized as follows. Section \ref{S2} describes the system model of grant-free random access. Section \ref{S3} introduces our proposed scheme and algorithms for JADD, and how SIC and channel coding are integrated into OAMP-MMV algorithms. In Section \ref{S4}, we analyze the performance of the proposed algorithms by focusing on the SE and computational complexity. Simulation results are given in Section \ref{S5} to demonstrate the superiority of our proposed algorithm. Finally, conclusions are drawn in Section \ref{S6}.

\textit{Notation}: The boldface lower and upper-case symbols denote column vectors and matrices, respectively. $(\cdot)^*$, $(\cdot)^{\rm T}$, $(\cdot)^{\rm H}$, $(\cdot)^{-1}$, ${\rm E}[\cdot]$ and ${\rm var}[\cdot]$ denote the conjugate, transpose, conjugate transpose, inversion, expectation and variance operators, respectively. $\|{\bf a}\|_p$ is the ${\ell_p}$ norm of vector ${\bf a}$. $\text{diag}({\bf a})$ is a diagonal matrix with the elements of vector ${\bf a}$ on its diagonal. ${\bf I}_N$ denotes the identity matrix of size $N\times N$, and ${\bf O}_{M\times N}$ denotes the null matrix of size ${M\times N}$. ${\bf 0}_N$ denotes the vector of size $N$ with all the elements being zeros, and $\O$ denotes the empty set. ${\bf A}_{m,:}$ and ${\bf A}_{:,n}$ are the $m$-th row vector and the $n$-th column vector of matrix ${\bf A}\in{\mathbb C}^{M\times N}$, respectively. ${\cal C}{\cal N}({{\bf x};\bm{\mu}},{\bf \Gamma})$ denotes the complex Gaussian distribution of a random vector ${\bf x}$ with mean vector $\bm{\mu}$ and covariance matrix ${\bf \Gamma}$. $\Re \mathfrak{e} \{\cdot\}$ denotes the real part of the corresponding arguments, and ${\rm card}\{\cdot\}$ denotes the cardinality of a set.

\vspace{-1mm}
\section{System Model}\label{S2}
We consider a typical uplink grant-free massive access scenario with mMTC, where OFDM is adopted to overcome the time dispersive channels. As shown in Fig. \ref{FIG1}(a), a single-antenna BS serves $K$ single-antenna IoT devices, where $K$ can be very large but only $K_a$ $(K_a\ll K)$ devices are active in each OFDM symbol. The data symbol $x_{k,t}$ sent by the $k$-th device during the $t$-th OFDM symbol belongs to a modulation constellation set $\Omega =\{a_1, a_2, \cdots, a_L\}$. If the device is silent in a period, then $x_{k,t}$ is equal to zero, i.e., no signal is transmitted. To distinguish the signals of the $k$-th device, $x_{k,t}$ is spread across $M$ subcarriers through a unique spreading sequence ${\bf s}_k \in \mathbb{C}^{M\times 1}$. At the BS, the received signal ${\bf y}_t \in \mathbb{C}^{M\times 1}$ can be written as
\begin{equation}\label{ytsum}
{\bf y}_t =\sum\limits_{k=1}^K {\bf H}_k {\bf s}_k \alpha_{k,t} x_{k,t} + {\bf w}_t,
\end{equation}
where ${\bf H}_k =\text{diag} \left([h_{1,k}, h_{2,k}, \cdots, h_{M,k}]^{\rm T}\right) \in \mathbb{C}^{M\times M}$, $h_{m,k}$ is the subchannel between the $k$-th device and the BS, $\alpha_{k,t}$ is the binary activity indicator that is equal to one if the $k$-th device is active and to zero otherwise, ${\bf w}_t$ is the additive white Gaussian noise (AWGN) that is distributed as ${\bf w}_t \sim{\cal CN}\left({\bf w}_t; {\bf 0}_M, \sigma^2{\bf I}_M \right)$. We assume that ${\rm E}\left[|x_{k,t}|^2\right] =1$ for active devices due to power control.

By defining ${\bf S} =\left[{\bf H}_1{\bf s}_1, {\bf H}_2{\bf s}_2, \cdots, {\bf H}_K{\bf s}_K \right] \in \mathbb{C}^{M\times K}$ and ${\bf x}_t =[\alpha_{1,t}x_{1,t}, \alpha_{2,t}x_{2,t}, \cdots, \alpha_{K,t}x_{K,t}]^{\rm T} \in \mathbb{C}^{K\times 1}$, then we have
\begin{equation}\label{yt}
{\bf y}_t ={\bf S}{\bf x}_t + {\bf w}_t.
\end{equation}

For mMTC, the number of potential IoT devices $K$ can be massively large but the number of radio resources for random access $M$ is limited, i.e., $K \gg M$. Therefore, it is impossible to employ traditional linear detectors, e.g., least-squares (LS) and linear minimum mean square error (LMMSE) detectors for solving ${\bf{x}}_t$ due to the under-determined problem in (\ref{yt}). On the other hand, due to their sporadic traffic, the number of active devices $K_a$ is much smaller than $K$, which indicates that the data vector ${{\bf{x}}_{t}}$ is sparse \cite{mMTC1,mMTC_new}.
In view of this fact, the JADD can be formulated as the following optimization problem
\begin{equation}\label{opt}
\begin{aligned}
&\ \ \ \ \ {\bf {\hat x}}_t =\mathop{\arg\min} \limits_{{\bf x}_t} \left\|{\bf x}_t\right\|_0, \\
&\mathrm{s.t.}\left\|{\bf y}_t - {\bf S}{\bf x}_t\right\|_2^2\le\varepsilon, x_{k,t} \in \{\Omega \cup 0 \}, \forall k,
\end{aligned}
\end{equation}
where the constant $\varepsilon$ is associated with the noise level of the system.

We can observe that the JADD in (\ref{opt}) is an SMV CS problem. Therefore, sparse signal recovery algorithms can be employed to solve (\ref{opt}). In the next section, we will present a slotted grant-free random access protocol, where the JADD can be further formulated as an MMV CS problem for achieving better massive access performance.

\vspace{-1.5mm}
\section{Proposed Scheme}\label{S3}

In this section, we propose a beacon-aided grant-free access scheme for mMTC. First, the frame structure and access procedure are presented. Then, we formulate the JADD problem as an MMV CS problem, whereby the inherently structured sparsity is exploited to design the OAMP-MMV-SSL algorithm and OAMP-MMV-ASL algorithm. Finally, SIC and channel coding are integrated into the OAMP-MMV algorithms for achieving better performance.

\vspace{-2mm}
\subsection{Frame Structure and Access Procedure}\label{S3.1}
\begin{figure}
\vspace*{-4mm}
\hspace{-3.5mm}
\centering
\subfloat[]{
\includegraphics[width=4.5 cm]
{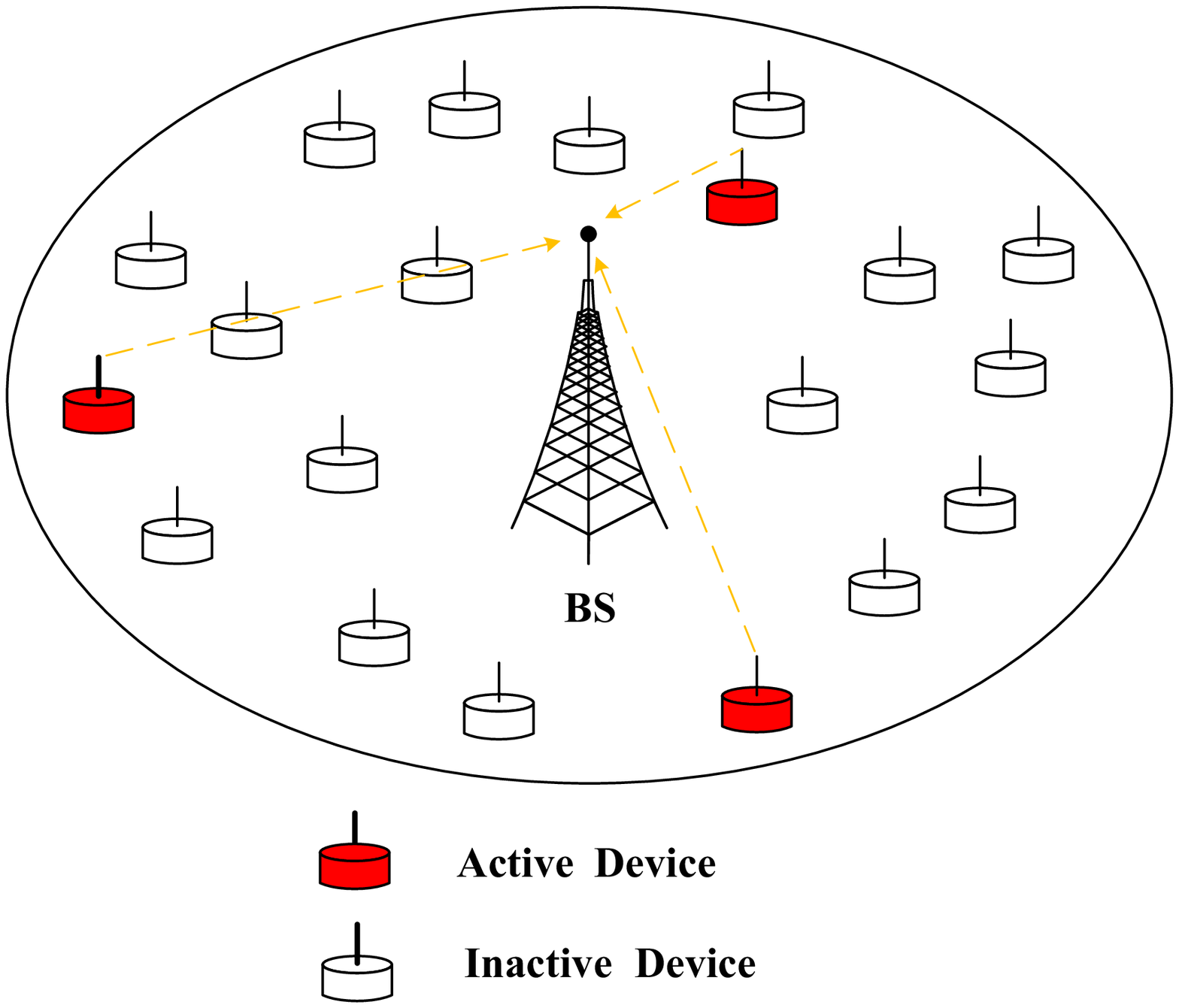}}
\hspace{7.5mm}
\subfloat[]{
\includegraphics[width=6 cm]
{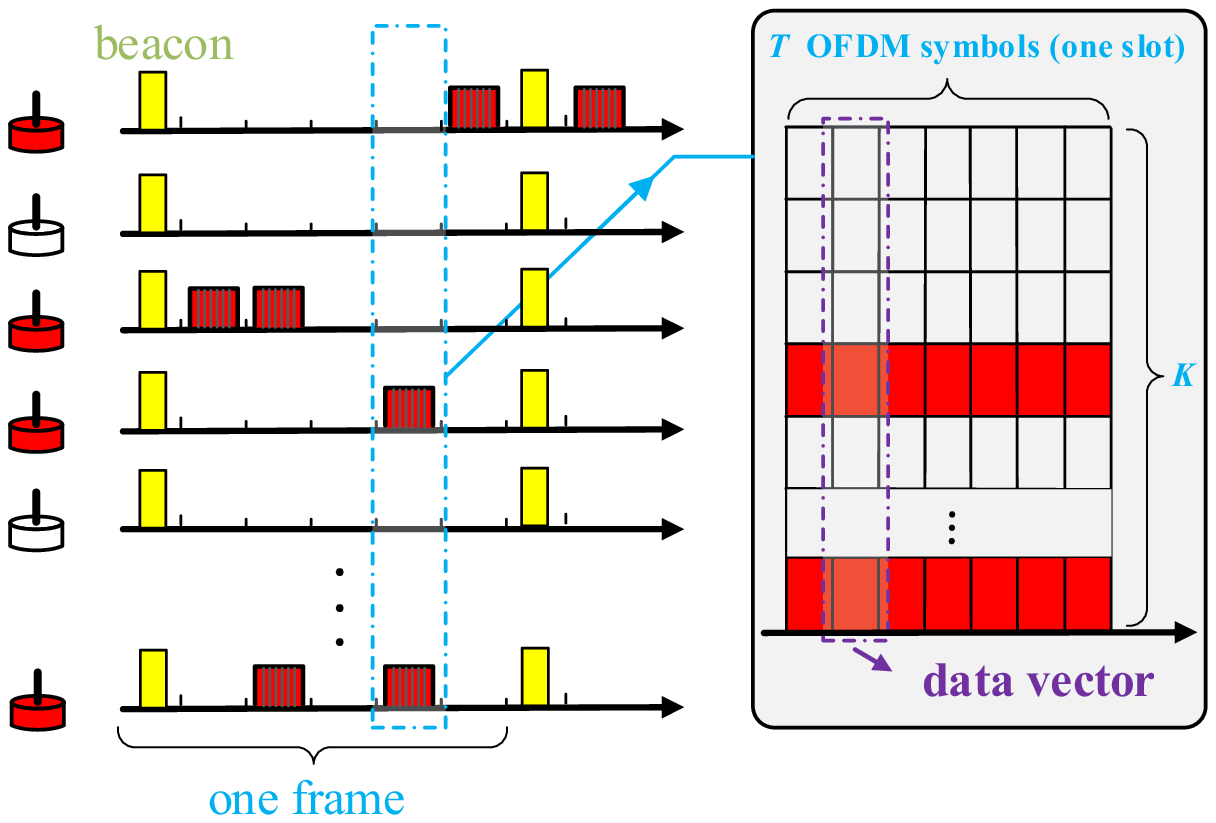}}
\captionsetup{font={footnotesize}, singlelinecheck = off, justification = raggedright,name={Fig.},labelsep=period}
\centering
\caption{(a) The activity of devices in mMTC exhibits sparsity; (b) Frame structure of the proposed beacon-aided grant-free access scheme, where the beacon is periodically broadcast by the BS to facilitate the synchronization, power control, and channel estimation at the IoT devices.}
\label{FIG1}
\vspace{-2mm}
\end{figure}
We consider that each frame starts with a beacon broadcast from the BS. The transmission of beacons is followed by the uplink random access phase, which consists of multiple time slots and each time slot consists of $T$ OFDM symbols, as shown in Fig. \ref{FIG1}(b). This periodic beacons can facilitate the IoT devices to achieve synchronization, power control, reliable channel estimation, etc.
We consider the accurate acquisition of CSI at the device, namely ${\bf H}_k{\bf {\hat H}}_k^{-1} ={\bf I}$, where ${\bf {\hat H}}_k = \text{diag}([{\hat h_{1,k}}, {\hat h_{2,k}}, \cdots, {\hat h_{M,k}}]^{\rm T}) \in \mathbb{C}^{M\times M}$ denotes the estimated channel. Previous works, e.g., \cite{JSAC_DY,CL_WBC1,CL_WBC2,TWC_DY,IORLS,TVT_SBL,CL_WC,TVT_Shim}, requires the CSI to be known at the BS. This indicates that extra pilot signals are necessary before the data transmission phase \cite{MA5B}, which increases the access latency. By contrast, the proposed scheme can fully exploit the inherent beacons that are periodically broadcast by the BS so that the random access procedure can be considerably simplified.

By performing the pre-equalization before transmitting the signals, the equivalent sensing matrix is
\vspace{-1mm}
\begin{equation}\label{stilde}
\begin{aligned}
{\bf {\widetilde S}}
& =\left[{\bf H}_1{\bf {\hat H}}_1^{-1}{\bf s}_1, {\bf H}_2{\bf {\hat H}}_2^{-1}{\bf s}_2, \cdots, {\bf H}_K{\bf {\hat H}}_K^{-1}{\bf s}_K\right] \\
& =\left[{\bf s}_1,{\bf s}_2, \cdots, {\bf s}_K\right].
\end{aligned}
\end{equation}

Accordingly, the received signal at the BS can be rewritten as
\vspace{-1mm}
\begin{equation}\label{yt_1}
{\bf y}_t ={\bf {\widetilde S}}{\bf x}_t +{\bf w}_t.
\end{equation}

By stacking the received signals in $T$ OFDM symbols (i.e., one time slot), we have
\vspace{-1mm}
\begin{equation}\label{Y}
{\bf Y} ={\bf {\widetilde S}}{\bf X} +{\bf W},
\end{equation}
where ${\bf Y} =\left[{\bf y}_1, {\bf y}_2, \cdots, {\bf y}_T\right]\in \mathbb{C}^{M\times T}$, ${\bf X} =[{\bf x}_1,{\bf x}_2, \cdots, {\bf x}_T]\in \mathbb{C}^{K\times T}$ and ${\bf W} =\left[{\bf w}_1,{\bf w}_2, \cdots, {\bf w}_T\right]$. As illustrated in Fig. \ref{FIG1}(b), the IoT devices' activity remains unchanged during $T$ OFDM symbols, i.e.,
\vspace{-1mm}
\begin{equation}\label{sparsitypattern}
{\rm{supp}}\left\{ {{{\bf{x}}_1}} \right\} = {\rm{supp}}\left\{ {{{\bf{x}}_2}} \right\} = \cdots = {\rm{supp}}\left\{ {{{\bf{x}}_T}} \right\},
\end{equation}
where ${\rm supp}\{\cdot\}$ denotes the set of non-zero elements in a vector. Given ${\bf {\widetilde S}}$ and ${\bf Y}$, the aim of JADD is to estimate the sparse matrix ${\bf X}$ and to identify the active devices, which results in an MMV CS problem. By leveraging the sparsity structure in (\ref{sparsitypattern}), a more accurate identification of the active devices and improved data detection performance can be obtained.

\vspace{-1.5mm}
\subsection{Spreading Code}\label{S3.2}

In conventional AMP algorithm, the sensing matrix is usually required to be a random matrix with i.i.d. elements, e.g., an i.i.d. Gaussian random matrix. Sensing matrices whose elements are not i.i.d. could deteriorate the performance of the AMP algorithm. On the other hand, the OAMP algorithm relaxes the sensing matrix requirement and can work well for partial unitary matrix\footnote{A matrix ${\bf A}\in \mathbb{C}^{M\times K}(M<K)$ is termed as partial unitary matrix if it satisfies ${\bf AA}^{\rm H}={\bf I}_M$.}. Therefore, it can be applied to a wider range of applications as compared with the AMP algorithm.

In this paper, we adopt a partial discrete Fourier transformation (DFT) matrix as the sensing matrix due to the following two reasons: (a) matrix inversion can be avoided, which reduces the computational complexity; (b) a partial DFT matrix is more rhythmic than Gaussian random matrix, which is beneficial for quantization and storage. The generation of the partial DFT matrix is performed as follows. Define ${\bf F}\in \mathbb{C}^{K\times K}$ as a DFT unitary matrix, i.e., ${\bf FF}^{\rm H}={\bf I}_K$, and ${\bf P}\in \mathbb{C}^{M\times K}$ as a selection matrix generated by randomly extracting $M$ rows from ${\bf I}_K$. Then, we have ${\bf {\widetilde S}}={\bf PF}$.

It is worth noting that the data symbol can be spread not only in the multiple frequency-domain subcarriers, but also in the multiple time-domain chips, if narrowband systems are considered. These two mathematical formulations are equivalent from the signal processing perspective. In this paper, we mainly consider the former methods that is used in wideband OFDM systems (without loss of generality).

\vspace{-1mm}
\subsection{OAMP-MMV Algorithm for JADD}\label{S3.3}
\begin{figure*}[!tp]
\vspace*{-4mm}
\hspace{-4mm}
\centering
\subfloat[]{
\begin{minipage}{0.5\textwidth}
\centering
\includegraphics[width=6 cm, keepaspectratio]
{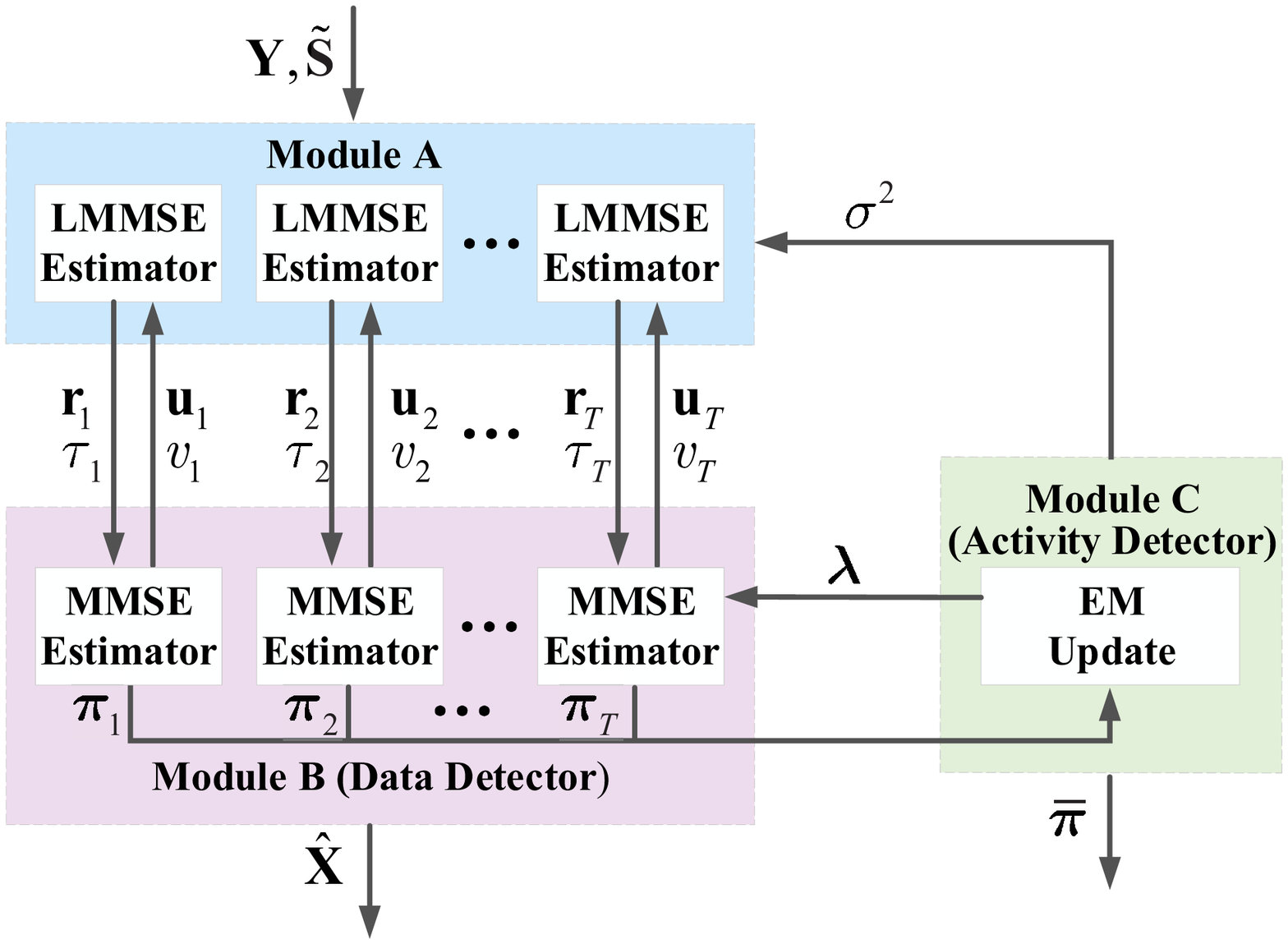}
\end{minipage}}
\subfloat[]{
\begin{minipage}{0.5\textwidth}
\centering
\includegraphics[width=6 cm, keepaspectratio]
{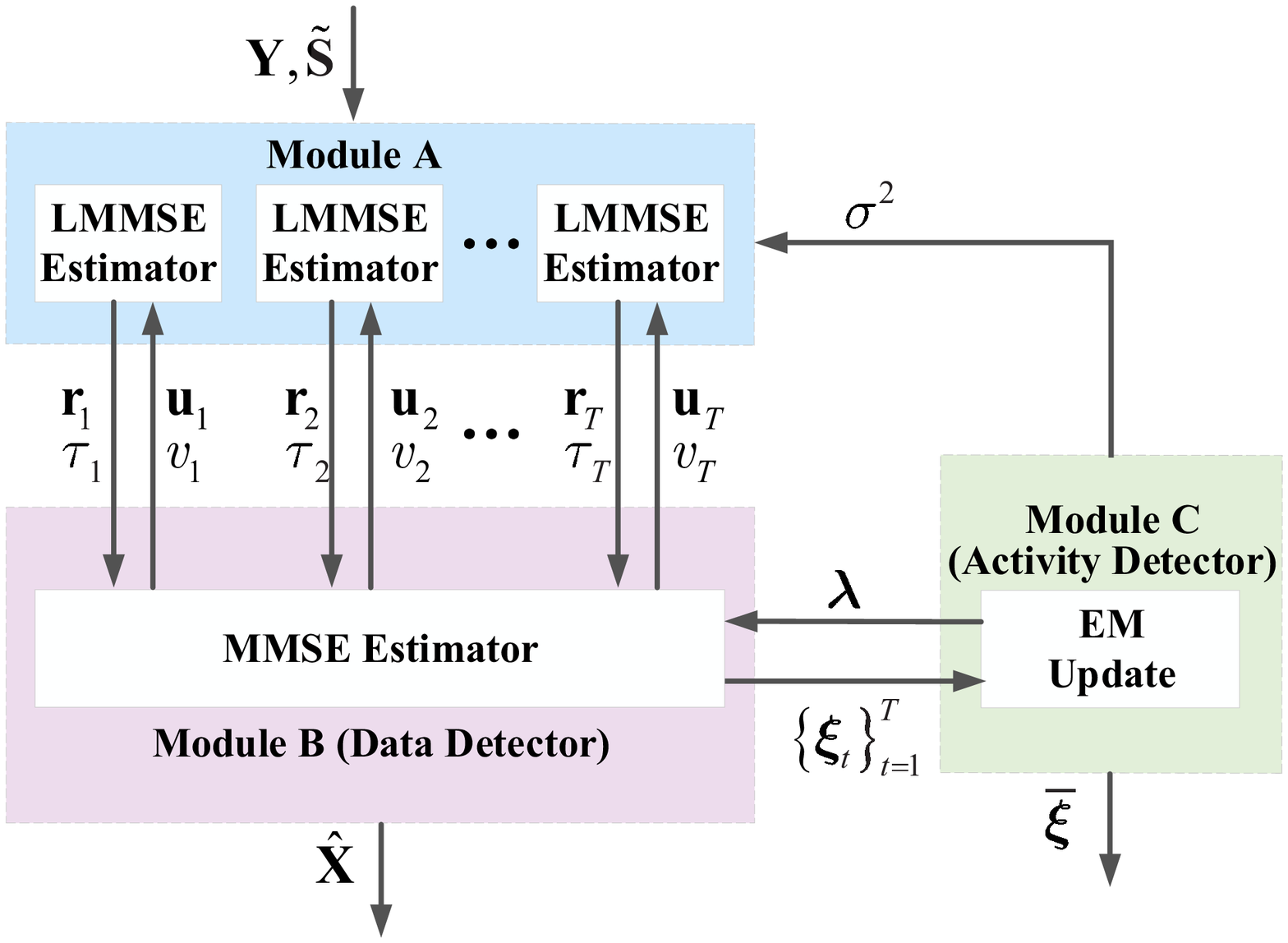}
\end{minipage}}
\captionsetup{font={footnotesize}, singlelinecheck = off, justification = raggedright,name={Fig.},labelsep=period}
\caption{Block diagram of the proposed (a) OAMP-MMV-SSL algorithm; (b) OAMP-MMV-ASL algorithm.}
\label{FIG2}
\vspace{-0mm}
\end{figure*}

Different from the conventional OAMP algorithm that limited to the SMV problem, we propose the OAMP-MMV-SSL algorithm and OAMP-MMV-ASL algorithm to exploit the structured sparsity in (\ref{sparsitypattern}). Particularly, the linear estimation (LE) module and non-linear estimation (NLE) module of OAMP algorithm are denoted as module A and module B, respectively, as shown in Fig. \ref{FIG2}. In addition, we add module C consisting of EM update. Specifically, the module B works as the data detector since it includes the minimum mean square error (MMSE) estimation of $\bf{X}$, and module C outputs the posterior sparsity ratio at the final stage for activity identification.
\subsubsection{OAMP-MMV-SSL Algorithm}\label{S3.3.1}
Previous works, e.g., \cite{JSAC_DY,CL_WBC1,CL_WBC2,TWC_DY}, fail to fully exploit the a priori distribution of random access signals. By contrast, our proposed algorithm accounts, in the a priori distribution, for the feature that access signals originate from a discrete constellation set. Specifically, the distribution of data symbol $x_{k,t}$ of the $k$-th device during the $t$-th OFDM symbol is assumed to be
\begin{equation}\label{prior}
p(x_{k,t}) =(1 -\lambda_{k,t})\delta(x_{k,t}) +\dfrac{\lambda_{k,t}}{L}\sum\limits_{l=1}^L \delta(x_{k,t}-a_l),
\end{equation}
where $\lambda_{k,t}\in[0,1]$ denotes the non-zero probability of $x_{k,t}$ and is termed as the sparsity ratio, and $\delta(\cdot)$ denotes the Dirac function.

In a conventional OAMP algorithm, the module A is used to decouple the received signal ${\bf Y}$. After some matrix manipulations, we have
\begin{equation}\label{LE}
{\bf r}_t^i ={\bf u}_t^{i-1} +\dfrac{K}{M}{\bf {\widetilde S}}^{\rm H}\left({\bf y}_t -{\bf {\widetilde S}}{\bf u}_t^{i-1}\right),
\end{equation}
where ${\bf r}_t^i$ is an intermediate variable that is output from module A, $i$ denotes the $i$-th iteration, and ${\bf u}_t$ is the NLE of ${\bf x}_t$, as described in (\ref{NLE}). We define $\tau_t^i =\frac{1}{K}{\rm E}\left[\left\|{\bf r}_t^i -{\bf x}_t\right\|_2^2\right]$ as the error measure of module A. On the other hand, module B is based on the assumption that ${{\bf{r}}_t}$ is modeled as an AWGN observation of the real signals ${\bf x}_t$, i.e.,
\vspace{-1mm}
\begin{equation}\label{NLE_base}
{\bf r}_t ={\bf x}_t +\sqrt{\tau_t}{\bf z}_t,
\end{equation}
where ${\bf z}_t\sim{\cal CN}\left({\bf z}_t; {\bf 0}_K, {\bf I}_K\right)$ and ${\bf z}_t$ is independent of ${\bf x}_t$. Then, the NLE of module B is calculated as
\vspace{-1mm}
\begin{equation}\label{NLE}
u_{k,t}^i =C_t^i\left({\rm E}\left[x_{k,t}|r_{k,t}^i\right]-\dfrac{\bar{\gamma}^i_t}{\tau_t^i}r_{k,t}^i\right),\ \ \ \forall k,
\end{equation}
where $C_t^i =\frac{\tau_t^i}{\tau_t^i - \bar{\gamma}^i_t}$ and $\bar{\gamma}^i_t =\frac{1}{K}\sum\limits_{k=1}^K \mathop{\rm var}\left[ x_{k,t}|r_{k,t}^i\right]$. The error measure of the NLE is $v_t^i =\frac{1}{K}{\rm E}\left[\left\|{\bf u}^i_t - {\bf x}_t\right\|_2^2\right]$.
These two modules are executed iteratively until convergence. Finally, the output of OAMP algorithm is ${\hat x}_{k,t}={\rm E}\left[x_{k,t}|r_{k,t}^i\right], \forall k$. In addition, $\tau_t$ and $v_t$ are defined as \cite{OAMP}
\begin{align}
\tau_t^i &=\frac{K-M}{M}v_t^{i-1} +\frac{K}{M}\sigma^2, \label{tao} \\
v_t^{i} &=\left(\dfrac{1}{\bar{\gamma}^i_t} -\dfrac{1}{\tau_t^i}\right)^{-1}. \label{v}
\end{align}

Combining the observation model (\ref{NLE_base}) with the a priori distribution in (\ref{prior}), we can obtain the approximate posterior distribution of ${{x}_{k,t}}$, which can be expressed as
\begin{equation}\label{posterior}
p(x_{k,t}|r^i_{k,t}) =(1-\pi^i_{k,t})\delta(x_{k,t}) +\pi^i_{k,t}\dfrac{\sum\limits_{l=1}^L q^i_{k,t,l} \delta(x_{k,t}-a_l)}{\sum\limits_{l=1}^L q^i_{k,t,l}},
\end{equation}
where
\begin{equation}\label{pi}
\pi^i_{k,t} =\left[1+(1-\lambda_{k,t})/(\frac{\lambda_{k,t}}{L}\sum\limits_{l=1}^L q^i_{k,t,l})\right]^{-1},
\end{equation}
and
\begin{equation}\label{q_l}
q^i_{k,t,l} = \exp \left( { - \frac{{{{\left| {{a_l}} \right|}^2} - 2{\Re} \mathfrak{e} \left\{ {{a_l^*}{r^i_{k,t}}} \right\}}}{{{\tau}_t^i}}} \right).
\end{equation}
The parameter $\pi_{k,t}\in[0,1]$ is termed as the posterior sparsity ratio, which represents the non-zero probability of $x_{k,t}$ based on the observation $r_{k,t}$. When $\pi_{k,t}\rightarrow0$, $p(x_{k,t}|r^i_{k,t})$ can be well approximated by a Dirac function, which implies that the $k$-th device is inactive in the $t$-th OFDM symbol. On the contrary, when ${\pi_{k,t}}\rightarrow 1$, the $k$-th device can be identified as an active device. Given (\ref{posterior}), the posterior mean and variance can be formulated as
\vspace{-2mm}
\begin{align}
\mu^i_{k,t} &={\rm E}\left[x_{k,t}|r_{k,t}^i\right] =({\pi^i_{k,t}\sum\limits_{l=1}^L a_lq^i_{k,t,l}})/({\sum\limits_{l=1}^L q^i_{k,t,l}}), \label{postmean} \\
\gamma^i_{k,t} &=\mathop{\rm var} \left[x_{k,t}|r_{k,t}^i\right] \nonumber \\ &=({\pi^i_{k,t}\sum\limits_{l=1}^L |a_l|^2q^i_{k,t,l}})/({\sum\limits_{l=1}^L q^i_{k,t,l}}) -|\mu^i_{k,t}|^2. \label{postvar}
\end{align}

Equations (\ref{LE})-(\ref{postvar}) constitute the key steps of the conventional OAMP algorithm \cite{OAMP}.
However, the conventional OAMP algorithm takes the noise variance $\sigma^2$ and sparsity ratio $\lambda$ as known parameters, which is not easy to obtain in general. To overcome this issue, we employ the EM algorithm \cite{EM} to learn the unknown parameters, e.g., $\bm{\theta} =\left\{ \sigma^2, \lambda_{k,t}, \forall k,t\right\}$. The EM algorithm consists of two steps
\begin{align}
Q\left(\bm{\theta}, \bm{\theta}^{i-1}\right) &= {\rm E}\left[\ln p\left({\bf X},{\bf Y}\right)|{\bf Y};\bm{\theta}^{i-1}\right], \label{E_step} \\
\bm{\theta}^{i} &= \arg \mathop{\max}\limits_{\bm{\theta}} Q\left(\bm{\theta}, \bm{\theta}^{i-1}\right), \label{M_step}
\end{align}
where ${\rm E}\left[(\cdot)|{\bf Y};\bm{\theta}^{i-1}\right]$ denotes the expectation conditioned on ${\bf Y}$ with parameters $\bm{\theta}^{i-1}$. The posterior distribution in (\ref{E_step}) can be approximately replaced by (\ref{posterior}) from the OAMP algorithm. However, the joint optimization of $\bm{\theta}$ can be difficult. Therefore, we adopt the incremental EM algorithm \cite{in_EM}, which updates one parameter in each iteration while keeping the other parameters fixed. By taking the partial derivative of (\ref{E_step}) with respect to each element of $\bm{\theta}$ and setting the derivatives equal to zero, we obtain the update rules of $\bm{\theta}$ as (see Appendix A for the detailed derivation)
\begin{align}
\lambda_{k,t}^{i} &=\pi_{k,t}^i, \ \ \ \forall k,t, \label{sparsity_up} \\
\left(\sigma^2\right)^{i} &= \dfrac{1}{T}\sum\limits_{t=1}^T \left[\dfrac{1}{M}\left\|{\bf y}_t-{\bf {\widetilde S}}\bm{\mu}_t^i \right\|_2^2 +\bar{\gamma}_t^i\right]. \label{noisevar_up}
\end{align}

Since the EM algorithm may converge to a local extremum or to a saddle point, a proper initialization is necessary. In our experiments, we find that the EM algorithm can work well when the initialization parameters change in a large range. Even so, the detection performance will deteriorate when the deviation between the initialization and the true value is too large. Therefore,
\color{black}following \cite{EM}, the initialization can be tailored as
\begin{align}
\lambda_{k,t}^0 &=\dfrac{M}{K}\mathop{\max}\limits_{c>0}\dfrac{1-2K\left[(1+c)^2\Phi(-c)-c\phi(c)\right]/M}{1+c^2-2\left[(1+c)^2\Phi(-c) -c\phi(c)\right]}, \forall k,t, \label{sparsity_int} \\
(\sigma^2)^0 &= \dfrac{1}{T}\sum\limits_{t=1}^T \dfrac{\left\|{\bf y}_t\right\|_2^2}{\left(\text{SNR}^0+1\right)M}, \label{noisevar_int}
\end{align}
where $\Phi(\cdot)$ and $\phi(\cdot)$ are the cumulative distribution function and probability distribution function of a standard Gaussian random variable, respectively, and $\text{SNR}^0$ is set to 100 as suggested in \cite{EM}. Equation (23) sets $\lambda_{k,t}^0$ to the theoretical phase transition point of $\ell_1$ norm optimization based sparse signal recovery, and the second equation can be verified by substituting the received signal ${\bf y}_t ={\bf S}{\bf x}_t + {\bf w}_t$ and the signal-noise-ratio (SNR) $= \|{\bf S}{\bf x}_t\|_2^2 / \|{\bf w}_t\|_2^2$ into the rightside. $\text{SNR}^0$ is an alternative since the true SNR is unknown. The numerical results show that the proposed schemes can work well when $\text{SNR}^0=100$.

\color{black}
Therefore, the OAMP algorithm is capable of solving the SMV problem in (\ref{yt}), in which the EM steps are integrated for learning the unknown parameters ${\bm{{\theta}}}$. If this approach is directly applied to the MMV problem in (\ref{Y}) and the transmitted signals ${\bf{X}}$ are estimated column by column, however, the sparsity structure is not fully exploited.
Therefore, in our proposed OAMP-MMV-SSL algorithm, we consider the following steps to learn the structured sparsity.
In particular, $\lambda_{k,t}$ represents the non-zero probability of $x_{k,t}$ and is independently updated in (\ref{sparsity_up}). In view of this fact, we can refine $\lambda_{k,t}$ as
%
\begin{equation}\label{sparsity_MMV}
\lambda _{k,1}^{i } = \cdots = \lambda _{k,T}^{i } = {{\bar{\pi}}^i_k} \triangleq \dfrac{1}{T}\sum\limits_{t = 1}^T {\pi _{k,t}^i}.
\end{equation}

Compared to (\ref{sparsity_up}), equation (\ref{sparsity_MMV}) learns the structured sparsity by averaging the posterior sparsity ratio. Then, this sparsity structure can be exploited by the MMSE estimator in the next iteration.
The EM update rules in (\ref{noisevar_up}) and (\ref{sparsity_MMV}) constitute the module C. We summarize the proposed OAMP-MMV-SSL algorithm in Algorithm \ref{Algorithm_OAMP}, whose flow chart diagram and procedure are illustrated in Fig. \ref{FIG2}(a).
\SetAlgoNoLine
\SetAlFnt{\small}
\SetAlCapFnt{\normalsize}
\SetAlCapNameFnt{\normalsize}\
\begin{algorithm}[!t]
\caption{OAMP-MMV-SSL Algorithm}\label{Algorithm_OAMP}
\begin{algorithmic}[1]
\REQUIRE Noisy observations ${\bf{Y}}$, spreading code matrix ${\bf{\widetilde S}}$, and the maximum number of iterations $I_{\mathrm{iter}}$.
\ENSURE Detected data matrix ${\hat{\bf{X}}} = [{{\hat{\bf{x}}}_1},{{\hat{\bf{x}}}_2},$ $ \cdots, {{\hat{\bf{x}}}_T}]$ and the posterior sparsity ratio $ {{\bar{\bm{\pi}}}}$.
\STATE ${\forall k,t}$: Set iteration index $i$ to $1$. Initialize ${{\bf{u}}_t^0 = \bm{0}_K}$ and ${{{{v_t^0}}}} =1$. ${{\lambda^{0}_{k,t}}}$ and $({\sigma^2})^{0}$ are initialized as (\ref{sparsity_int}) and (\ref{noisevar_int}), respectively.
\FOR {$ i=1$ to $ I_{\mathrm{iter}}$}
\STATE \textbf{\textbf{\%}Module A:}
\STATE ${\forall t}$: $
{{\bf{r}}_t^{i}} = {{\bf{u}}_t^{i-1}} + \frac{K}{M}{{\bf{\widetilde S}}^{\rm{H}}}\left( {{{\bf{y}}_t} - {\bf{\widetilde S}}{{\bf{u}}_t^{i-1}}} \right)$, \ \ \ ${\tau}_t^{i} = \frac{K-M}{M}{v}_t^{i-1} + \frac{K}{M}({\sigma^2})^{i-1}$.
\STATE \textbf{\textbf{\%}Module B:}
\STATE ${\forall k,t,l}$: $q^i_{k,t,l} = \exp \left( { - \frac{{{{\left| {{a_l}} \right|}^2} - 2{\Re} \mathfrak{e} \left\{ {{a_l^*}{r^i_{k,t}}} \right\}}}{{{\tau}_t^i}}} \right)$.
\STATE ${\forall k,t}$: ${\pi^i_{k,t}} = \left[{1+(1-{\lambda_{k,t}^{i-1}})/ (\frac{{\lambda_{k,t}^{i-1}}}{L}\sum\limits_{l = 1}^L {q^i_{k,t,l}}})\right]^{-1}$.
\STATE ${\forall k,t}$: ${\mu^i_{k,t}}= ({{{\pi^i _{k,t}}}}\sum\limits_{l = 1}^L {{a_l}q^i_{k,t,l}})/(\sum\limits_{l = 1}^L {q^i_{k,t,l}})$, \ \ \ ${\gamma^i_{k,t}} = ({{{\pi^i _{k,t}}}}\sum\limits_{l = 1}^L {|{a_l}|^2q^i_{k,t,l}})/(\sum\limits_{l = 1}^L {q^i_{k,t,l}})- {| {\mu^i_{k,t}}|^2}$.
\STATE ${\forall t}$: ${\bar{\gamma}^i_{t}}= \frac{1}{K}\sum\limits_{k = 1}^K \gamma_{k,t}^i$, \ \ \ ${C_t^i} = \frac{{{{ {{\tau}}
}_t^i}}}{{{{ {{\tau}}}_t^i} - {\bar{\gamma}^i_{t}}}}$, \ \ \ ${ {{v}}_t^{i}} = {\left( {\dfrac{1}{{\bar{\gamma}^i_{t}}} - \dfrac{1}{{{{{{\tau}}_t^{i}}}}}} \right)^{ - 1}}$.
\STATE ${\forall k,t}$: $u_{k,t}^i = {C_t^i}\left( { {\mu^i_{k,t}}- \frac{{\bar{\gamma}^i_{t}}}{{{{ {{\tau}}}_t^i}}}r_{k,t}^i} \right)$.
\STATE \textbf{\textbf{\%}Module C:}
\STATE ${\forall k,t}$: $\lambda _{k,t}^{i} = {{\bar{\pi}}^i_k} = \frac{1}{T}\sum\limits_{t = 1}^T {\pi _{k,t}^i}$.
\STATE ${\left( {\sigma^2} \right)^{i}} = \frac{1}{T}\sum\limits_{t = 1}^T \left[ {\frac{1}{M}{{ {{{\left\|{{\bf{y}}_t} - {\bf{\widetilde S}}{\bm{\mu}_t^i} \right\|}_2^2} + {{{\bar{\gamma}}_{t}^i}}}}}}\right]$.
\ENDFOR
\STATE ${\forall t}$: ${\hat{{\bf{x}}}_t} = {{\bm{\mu}}_{t}^i}$.
\end{algorithmic}
\end{algorithm}

\begin{figure*}[!tp]
\vspace*{-10mm}
\centering
\hspace*{0mm}
\subfloat[]{
\begin{minipage}{0.35\textwidth}
\includegraphics[width=6.5 cm, keepaspectratio]
{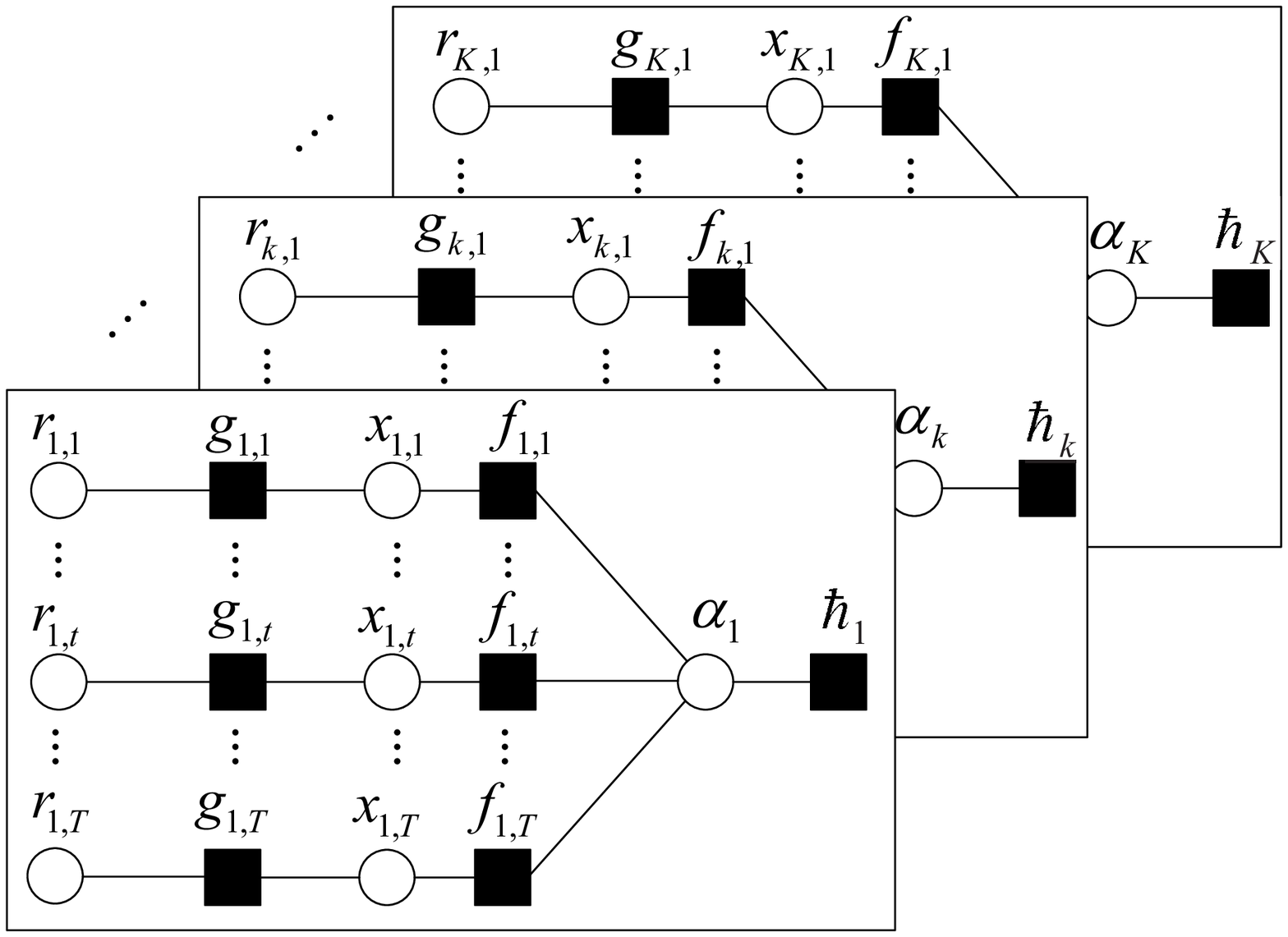}
\end{minipage}}
\hspace*{12mm}
\subfloat[]{
\begin{minipage}{0.52\textwidth}
\small
\begin{tabular}{c|c|c}
\Xhline{1pt}
Factor & Distribution & Function \\
\Xhline{1pt}
$g_{k,t}$ & $p(r_{k,t}|x_{k,t})$ & ${\cal{CN}}{(x_{k,t};r_{k,t},\tau_t)}$ \\
\Xhline{1pt}
$f_{k,t}$ & $p(x_{k,t}|\alpha_{k})$ & $\left( {1 - {\alpha_k}} \right)\delta \left( {{x_{k,t}}} \right) + \dfrac{{{\alpha_k}}}{L}\sum\limits_{l = 1}^L {\delta \left( {{x_{k,t}} - {a_l}} \right)} $\\
\Xhline{1pt}
$\hbar_{k}$ & $p(\alpha_k)$ & $\left( {1 - {\lambda _k}} \right)\delta \left( {{\alpha _k}} \right) + {\lambda _k}\delta \left( {{\alpha _k} - 1} \right)$ \\
\Xhline{1pt}
\end{tabular}
\end{minipage}}
\captionsetup{font={footnotesize}, singlelinecheck = off, justification = raggedright,name={Fig.},labelsep=period}
\caption{(a) Factor graph ${\cal G}$ for the joint distribution $p\left( {{\bf{X}},{\bf{R}},{\bm{\alpha}}} \right)$ in (\ref{MP_joint}); (b) The functions of the factor node in graph $\cal G$.}
\label{fp}
\vspace{-2mm}
\end{figure*}
Based on the output of Algorithm 1, we can identify the active devices and detect the corresponding data. Hence, the design of the activity detector is essential to the JADD performance. Specifically, the binary hypothesis testing of the activity detection can be written as follows
\begin{equation}\label{hypothesis_test}
\left\{ {\begin{array}{*{20}{cl}}
{{\cal H}_1:}&{{{ \hat{\alpha}}_{k,t}} = 1, \text{active}},\\
{{\cal H}_0:}&{{{ \hat{\alpha}}_{k,t}} = 0, \text{inactive}}.
\end{array}} \right.
\end{equation}
The corresponding two statuses are both contained in the posterior distribution in (\ref{posterior}). By integrating over the corresponding hypothesis space, we obtain the probability of ${\cal H}_1$ and ${\cal H}_0$
\begin{subequations} \label {p_hypothesis:1}
\begin{align}
& \Pr \left( {{\hat{\alpha} _{k,t}} = 1|{r_{k,t}}} \right) = \int\limits_{{{\cal H}_1}} {p\left( {{x_{k,t}}|{r_{k,t}}} \right)} d{x_{k,t}} = {\pi _{k,t}}, \label{p_hypothesis:1A} \\
& \Pr \left( {{\hat{\alpha} _{k,t}} = 0|{r_{k,t}}} \right) = 1 - {\pi _{k,t}}, \label{p_hypothesis:1B}
\end{align}
\end{subequations}
respectively. Then, a Bayesian activity detector can be formulated as
\begin{equation}\label{detector}
\dfrac{{\Pr \left( {{\hat{\alpha}_{k,t}} = 1|{r_{k,t}}} \right)}}{{\Pr \left( {{\hat{\alpha} _{k,t}} = 0|{r_{k,t}}} \right)}}
\mathop \gtrless \limits_{{\cal H}_0}^{{\cal H}_1} 1.
\end{equation}
Substituting (\ref{p_hypothesis:1}) into (\ref{detector}), we have
\begin{equation}\label{threshold}
\pi_{k,t}\mathop \gtrless \limits_{{\cal H}_0}^{{\cal H}_1} \dfrac{1}{2}, \ \ \ \forall k,t.
\end{equation}
Since a device is either active or inactive within a slot, the final activity detector is
\vspace{-1mm}
\begin{equation}\label{final_detector}
\bar \pi_k \mathop \gtrless \limits_{{\cal H}_0}^{{\cal H}_1} \dfrac{1}{2}, \ \ \ \forall k.
\end{equation}

\subsubsection{OAMP-MMV-ASL Algorithm}\label{S3.3.2}
In the OAMP-MMV-SSL algorithm, the a priori distributions of $\{{x_{k,t}}\}_{t=1}^T$ are assumed to be mutually independent. This simplifies the derivation of the posterior distribution and the MMSE estimation. However, this assumption fails to accurately characterize the true sparsity structure of random access signals. To this end, the OAMP-MMV-ASL algorithm is proposed by fully exploiting the sparsity ratio shared by $T$ different data symbols of the $k$-th device. More specifically, different from (\ref{prior}) that considers only the a priori distribution of $x_{k,t}$, the a priori distribution of ${{\bf{x}}_k} = [x_{k,1},x_{k,2},\cdots,x_{k,T}]^{\rm{T}}$ is jointly considered as
\vspace{-1mm}
\begin{align}
p\left( {{{\bf{x}}_k}|{\alpha_k}} \right)
&= \prod\limits_{t = 1}^T p\left( {{{{x}}_{k,t}}|{\alpha_k}} \right) \nonumber \\
&= \prod\limits_{t = 1}^T \left[ {\left( {1 - {\alpha_k}} \right)} \delta \left( {{x_{k,t}}} \right) + \dfrac{{{\alpha _k}}}{L}\sum\limits_{l = 1}^L {\delta \left( {{x_{k,t}} - {a_l}} \right)} \right]. \label{MP_x_prior}
\end{align}
Furthermore, the probability distribution of activity indicator ${\bm{\alpha}}=\left[\alpha_1,\alpha_2,\cdots,\alpha_K \right]^{\rm{T}}$ can be written as
\vspace{-1mm}
\begin{equation}\label{MP_lamda_prior}
p\left( {{{\bm{\alpha}}}} \right)
= \prod\limits_{k = 1}^K p\left( {{\alpha_k}} \right)
= \prod\limits_{k = 1}^K \left[ \left( {1 - {\lambda _k}} \right)\delta \left( {{\alpha _k}} \right) + {\lambda _k}\delta \left( {{\alpha _k} - 1} \right) \right],
\end{equation}
where ${\lambda_k}$ denotes the probability that ${\alpha_k} = 1$.

Based on the a priori distributions in (\ref{MP_x_prior})-(\ref{MP_lamda_prior}) and the AWGN observation model (\ref{NLE_base}), we obtain the joint probability distribution as follow
\begin{equation}\label{MP_joint}
\begin{aligned}
p\left( {{\bf{X}},{\bf{R}},{\bm {\alpha}}} \right)
& = p\left( {{\bf{R}}|{\bf{X}}} \right)p\left( {{\bf{X}}|{\bm{\alpha}}} \right)p\left( {\bm {\alpha}} \right) \\
& = \prod\limits_{k = 1}^K p\left( {{{\bf{r}}_k}|{{\bf{x}}_k}} \right)p\left( {{{\bf{x}}_k}|{{\alpha}_k}} \right)p\left( {{\alpha}_k} \right)  \\
& = \prod\limits_{k = 1}^K \underbrace {p\left( {{\alpha _k}} \right)}_{\hbar_{k}}{\prod\limits_{t = 1}^T \underbrace {p \left( {{r_{k,t}}|{x_{k,t}}} \right)}_{g_{k,t}}\underbrace {p\left( {{x_{k,t}}|{\alpha _k}} \right)}_{f_{k,t}}},
\end{aligned}
\end{equation}
where ${\bf{R}}=[{{\bf{r}}_1}, {{\bf{r}}_2}, \cdots, {{\bf{r}}_T}]{ \in \mathbb{C}^{K \times T}}$ and ${{\bf{r}}_k} = [r_{k,1},r_{k,2},\cdots,r_{k,T}]^{\rm{T}}{ \in \mathbb{C}^{T \times 1}}$. According to (\ref{MP_joint}), the corresponding factor graph ${\cal G}$ is illustrated in Fig. \ref{fp}(a), where the circles represent variable nodes, the squares represent factor nodes, and the meaning of the factor nodes is listed in Fig. \ref{fp}(b).

The goal of the proposed OAMP-MMV-ASL algorithm is to obtain the approximate posterior distribution $p({x_{k,t}}|{{\bf{r}}_{k}})$ for the MMSE estimation ${\hat{x}}_{k,t} = {\rm{ E}}\left[ {{x_{k,t}}|{\bf{r}}_{k}} \right]$. To this end, we apply the sum-product algorithm \cite{SP} to the graph ${\cal G}$ and obtain $p({x_{k,t}}|{{\bf{r}}_{k}})$ through message passing. The details of the derivation are elaborated in Appendix B. In particular, the $i$-th iteration of the proposed OAMP-MMV-ASL algorithm can be summarized as follows. First, $ {\bf{R}}^i$ and $\{{\tau}^i_t\}_{t=1}^T$ are obtained from module A. The message passing steps in module B are given in (\ref{m_ftol})-(\ref{m_post}) (see Appendix B). On this basis, the approximate posterior distribution of $x_{k,t}$ can be written as
\vspace{-1mm}
\begin{equation}\label{MP_posterior}
 p\left( {{x_{k,t}}|{{\bf{r}}^i_{k}}} \right) = \left( {1 - {\pi^{\prime,i}_{k,t}}} \right)\delta \left( {{x_{k,t}}} \right) + {{{\pi ^{\prime,i}_{k,t}}}}\dfrac{\sum\limits_{l = 1}^L {q^i_{k,t,l}} \delta \left( {{x_{k,t}} - {a_l}} \right)}{{\sum\limits_{l = 1}^L {q^i_{k,t,l}}}},
\end{equation}
where
\vspace{-1mm}
\begin{equation}\label{pi2}
{\pi^{\prime,i} _{k,t}} = \left[{1+(1-{\xi^i_{k,t}})/ (\dfrac{{\xi^i_{k,t}}}{L}\sum\limits_{l = 1}^L {q^i_{k,t,l}}})\right]^{-1},
\end{equation}
${\xi^i_{k,t}}$ is given in (\ref{kexi}), and $\pi^{\prime,i}_{k,t}$ denotes the posterior sparsity ratio. The structure of $p\left( {{x_{k,t}}|{{\bf{r}}_{k}}} \right)$ is similar to (\ref{posterior}) for the OAMP-MMV-SSL algorithm. However, the calculations of $p(x_{k,t}|{\bf{r}}_k)$ for the SSL and ASL models are different in two main aspects. As for the ASL model, the parameter ${\lambda}_{k,t}$ in (\ref{pi}) is replaced by ${\xi}_{k,t}$ in order to fully exploit the global information of ${\bf{r}}_k$ through message passing. Also, the posterior mean ${{\mu}_{k,t}^i}$ and variance ${{\gamma}_{k,t}^i}$ can still be calculated in (\ref{postmean}) and (\ref{postvar}), but ${\pi}_{k,t}$ needs to be replaced by ${\pi}^{\prime}_{k,t}$ in (\ref{pi2}). Moreover, similar to the OAMP-MMV-SSL algorithm, we adopt the EM algorithm to learn the unknown sparsity ratio $\lambda_k$ and noise variance $\sigma^2$ in module C. The update rule for $\lambda_k$ is
\begin{equation}\label{rou}
\lambda^{i}_{k} = {{\bar{\xi}}^i_k} \triangleq \dfrac{1}{T}\sum\limits_{t = 1}^T {\xi_{k,t}^{i}},
\end{equation}
and $(\sigma^2)^{i}$ is updated by employing (\ref{noisevar_up}). The OAMP-MMV-ASL algorithm is summarized in Algorithm \ref{Algorithm_OAMP_MP}.
\SetAlgoNoLine
\SetAlFnt{\small}
\SetAlCapFnt{\normalsize}
\SetAlCapNameFnt{\normalsize}
\begin{algorithm}[!t]
\caption{OAMP-MMV-ASL Algorithm}\label{Algorithm_OAMP_MP}
\begin{algorithmic}[1]
\REQUIRE ~ Noisy observation ${\bf{Y}}$, spreading code matrix ${\bf{\widetilde S}}$, and the maximum number of iterations $I_{\mathrm{iter}}$.
\ENSURE Detected data matrix ${\hat{\bf{X}}} = [{{\hat{\bf{x}}}_1},{{\hat{\bf{x}}}_2},$ $ \cdots, {{\hat{\bf{x}}}_T} ]$ and the updated sparsity ratio $ {{\bar{\bm{\xi}}}} $.
\STATE ${\forall k,t}$: Set iteration index $i$ to 1. Initialize ${{\bf{u}}_t^0 = \bm{0}_K}$ and ${{{{v_t^0}}}} =1$.
${{\lambda^{0}_{k}}}$ and $({\sigma^2})^{0}$ are initialized as (\ref{sparsity_int}) and (\ref{noisevar_int}), respectively.
\FOR {$ i=1 $ to $ I_{\mathrm{iter}} $}
\STATE \textbf{\textbf{\%}Module A:}
\STATE ${\forall t}$: ${{\bf{r}}_t^{i}} = {{\bf{u}}_t^{i-1}} + \frac{K}{M}{{\bf{\widetilde S}}^{\rm{H}}}\left( {{{\bf{y}}_t} - {\bf{\widetilde S}}{{\bf{u}}_t^{i-1}}} \right)$, \ \ \ ${\tau}_t^{i} = \frac{K-M}{M}{v}_t^{i-1} + \frac{K}{M}({\sigma^2})_t^{i-1}$.
\STATE \textbf{\textbf{\%}Module B:}
\STATE ${\forall k,t,l}$: $q^i_{k,t,l} = \exp \left( { - \frac{{{{\left| {{a_l}} \right|}^2} - 2{\Re} \mathfrak{e} \left\{ {{a_l^*}{r^i_{k,t}}} \right\}}}{{{\tau}_t^i}}} \right)$.
\STATE $\forall k,t$: ${\eta^i_{k,t}} = 1 - \frac{1}{{1 + \frac{1}{L}\sum\limits_{l = 1}^L {q^i_{k,t,l}}}}$, \ \ \ ${\xi^i_{k,t}} = \frac{{\lambda _k^{i-1}\prod\limits_{t' \in \backslash t} {\eta _{k,t'}^i}}}{{\left( {1 - \lambda _k^{i-1}} \right)\prod\limits_{t' \in \backslash t} {\left( {1 - \eta _{k,t'}^i} \right)} + \lambda _k^{i-1}\prod\limits_{t' \in \backslash t} {\eta _{k,t'}^i}}}$.
\STATE ${\forall k,t}$: ${\pi^{\prime,i}_{k,t}} = \left[{1+(1-{\xi_{k,t}^{i}})/ (\frac{{\xi_{k,t}^{i}}}{L}\sum\limits_{l = 1}^L {q^i_{k,t,l}}})\right]^{-1}$.
\STATE ${\forall k,t}$: ${\mu^i_{k,t}}= ({{{\pi^{\prime,i} _{k,t}}}}\sum\limits_{l = 1}^L {{a_l}q^i_{k,t,l}})/(\sum\limits_{l = 1}^L {q^i_{k,t,l}})$, \ \ \ ${\gamma^i_{k,t}} = ({{{\pi^{\prime,i} _{k,t}}}}\sum\limits_{l = 1}^L {|{a_l}|^2q^i_{k,t,l}})/(\sum\limits_{l = 1}^L {q^i_{k,t,l}})- {| {\mu^i_{k,t}}|^2}$.
\STATE ${\forall t}$: ${\bar{\gamma}^i_{t}}= \frac{1}{K}\sum\limits_{k = 1}^K \gamma_{k,t}^i$, \ \ \ ${C_t^i} = \frac{{{{ {{\tau}}}_t^i}}}{{{{ {{\tau}}}_t^i} - {\bar{\gamma}^i_{t}}}}$, \ \ \ ${ {{v}}_t^{i}} = {\left( {\dfrac{1}{{\bar{\gamma}^i_{t}}} - \dfrac{1}{{{{{{\tau}}_t^{i}}}}}} \right)^{ - 1}}$.
\STATE ${\forall k,t}$: $u_{k,t}^i = {C_t^i}\left( { {\mu^i_{k,t}}- \frac{{\bar{\gamma}^i_{t}}}{{{{ {{\tau}}}_t^i}}}r_{k,t}^i} \right)$.
\STATE \textbf{\textbf{\%}Module C:}
\STATE $\lambda^{i}_{k} = {{\bar{\xi}}^i_k} = \dfrac{1}{T}\sum\limits_{t = 1}^T {\xi_{k,t}^{i}}$.
\STATE ${\left( {\sigma^2} \right)^{i}} = \frac{1}{T}\sum\limits_{t = 1}^T \left[ {\frac{1}{M}{{ {{{\left\|{{\bf{y}}_t} - {\bf{\widetilde S}}{\bm{\mu}_t^i} \right\|}_2^2} + {{{\bar{\gamma}}_{t}^i}} }} }}\right]$.
\ENDFOR
\STATE $\forall t$, ${\hat{{\bf{x}}}_t} = {{\bm{\mu}}_{t}^i}$.
\end{algorithmic}
\end{algorithm}

In a nutshell, the main difference between the SSL and ASL models lies in the MMSE estimator (module B), as shown in Fig. \ref{FIG2}. It originates from the different formulations for the a priori distribution of the random access matrix $\bf{X}$. As for the SSL model, in particular, we assume the mutual independence of $\{{x_{k,t}}\}_{t=1}^T$ and exploit only the sparsity structure in the EM update rules. In contrast, the a priori distribution of the ASL model is closer to true distribution and is more accurate. By resorting to the sum-product rule, the calculation of ${\xi}_{k,t}$ exploits the whole information from ${\bf{r}}_k$. Except for the different sparsity exploitation methods, the EM update rules are the same for the ASL and SSL models. In the simulation section, we will compare the proposed OAMP-MMV-SSL and OAMP-MMV-ASL algorithms by considering different system configurations, and their advantages and limitations are discussed. We will also see that even if the system is slightly overloading (i.e., $M < K_a$), the OAMP-MMV-ASL algorithm can achieve considerable detection performance, as long as $T$ is large enough.

\subsection{SIC-Based JADD}\label{S3.4}
In this section, we introduce SIC methods to enhance the performance of the proposed slotted grant-free massive access scheme. SIC methods are useful in non-linear detection and provide a trade-off between computational complexity and reliability. SIC methods are beneficial for sparse signal recovery algorithms as well \cite{{SIC1},{SIC2}}. Taking problem (\ref{yt_1}) as an example, when the signal ${{{\bf{s}}_k}{\alpha _{k,t}}{x_{k,t}}} $ of the active devices is detected correctly and subtracted from the received signal ${{\bf{{y}}}_t}$, the residual ${{\bf{\widetilde{y}}}_t}$ is made of fewer active devices. In particular, the residual ${{\bf{\widetilde{y}}}_t}$ can be written as
\begin{equation}\label{y_sic_res}
{{\bf{\widetilde{y}}}_t} = \sum\limits_{k = 1, k \not\in \kappa}^K {{{\bf{s}}_k}{\alpha _{k,t}}{x_{k,t}}} + {{\bf{w}}_t},
\end{equation}
where $\kappa$ denotes the set of detected active devices. Therefore, the data vector becomes sparser after performing interference cancellation. After multiple SIC operations, the number of residual active devices becomes small, leading to improved detection performance.

However, reaping these advantages by using SIC methods faces several challenges, which include increasing the computational complexity, the associated error propagation, and the choice of criterion of cancellation, etc. Specifically, SIC-based detection algorithms are executed repeatedly, which dramatically increases the computational complexity. An intuitive solution to this problem is to limit the number of interference cancellations, which motivates us to subtract the signals of multiple active devices simultaneously instead of one active devices at each iteration. In this way, we can strive a balance between the system performance and the computational complexity. In addition, error propagation originates from the cumulative effects of the subtracted signals. To mitigate this error, we integrate channel coding into the SIC scheme, which increases the reliability of data detection thanks to the error correction capability of channel coding.
\begin{figure}[!tp]
\vspace*{-5mm}
\hspace*{3.5mm}
\centering
\includegraphics[width=4.95 cm,angle=270,keepaspectratio]
{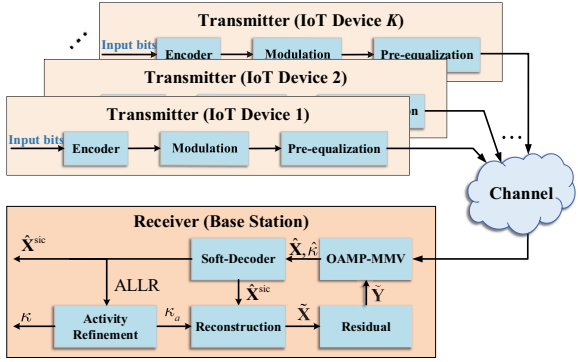}
\captionsetup{font={footnotesize}, singlelinecheck = off, justification = raggedright,name={Fig.},labelsep=period}
\caption{Block diagram of the proposed SIC-based JADD.}
\label{SICB}
\vspace*{-2mm}
\end{figure}

Furthermore, the cancellation ordering of SIC impacts the detection performance. This issue is usually tackled by subtracting the strongest signals first. To this end, the SNR or the signal-to-interference-and-noise-ratio (SINR) is usually used as the metrics for performing the ordering \cite{SIC_SNR}. However, due to power control with ${\rm{E}}\left[|x_{k,t}|^2\right] = 1$, the SNRs associated with different devices are almost the same. Also, the SINR is usually difficult to calculate in practice. Hence, we consider the LLR obtained in the soft decision. More specifically, the LLR is the logarithm of the ratio of the probability of transmitting a $0$ bit versus the probability of transmitting a $1$ bit for a received signal. For a specific bit $b$, the LLR is defined as
\begin{align}\label{LLR}
{{\Lambda}_{k,t,b}}
 &= \ln \left[ {\frac{{\Pr \left( {b = 0|{\hat{x}_{k,t}}} \right)}}{{\Pr \left( {b = 1|{\hat{x}_{k,t}}} \right)}}} \right] \nonumber \\
 &= \ln \left\{ {\frac{{\sum\limits_{a \in {\Omega _0}} {\exp \left\{ -{\frac{1}{{{\sigma ^2}}}\left[ {{{\left( {{x_0} - {a_x}} \right)}^2} + {{\left( {{y_0} - {a_y}} \right)}^2}} \right]} \right\}} }}{{\sum\limits_{a \in {\Omega _1}} {\exp \left\{ -{\frac{1}{{{\sigma ^2}}}\left[ {{{\left( {{x_0} - {a_x}} \right)}^2} + {{\left( {{y_0} - {a_y}} \right)}^2}} \right]} \right\}} }}} \right\},
\end{align}
where $b$ is the $b$-th bit position of $\hat{x}_{k,t}$, ${\Omega _0}$ and ${\Omega _1}$ are the sets of ideal constellation symbols with bit $0$ and $1$ in the $b$-th bit position, respectively, $(x_0,y_0)$ and $(a_x,a_y)$ are the coordinates of $\hat{x}_{k,t}$ and the ideal symbol $a$ in the constellation diagram, respectively. The noise variance $\sigma^2$ in (\ref{LLR}) can be obtained from module C of our proposed OAMP-MMV algorithm. Therefore, the signals whose module of LLR is larger are more likely to be correctly detected and decoded. Although the exact LLR can provide the greatest accuracy, obtaining such information is computationally expensive. Hence, as a compromise approach, we consider to employ the approximate LLR (ALLR), by trading the accuracy for higher efficiency, that is defined as
\begin{align}\label{Approximate LLR}
{{{\Lambda}}_{k,t,b}} \approx
&- \frac{1}{{{\sigma ^2}}}\left\{{ \mathop {\min }\limits_{a \in {\Omega _0}} \left[ {{{\left( {x_0 - {a_x}} \right)}^2} + {{\left( {y_0 - {a_y}} \right)}^2}} \right]} \right. \nonumber \\
&\left.{- \mathop {\min }\limits_{a \in {\Omega _1}} \left[ {{{\left( {x_0 - {a_x}} \right)}^2} + {{\left( {y_0 - {a_y}} \right)}^2}} \right]}\right\}.
\end{align}

To evaluate the reliability of the detected signal associated with the $k$-th device, we define the average ALLR metric as
\begin{equation}\label{Approximate LLR_mean}
{\overline{{\Lambda}}_{k}} = {\dfrac{1}{T{{{\log}_2}L}}\sum\limits_{t = 1}^T\sum\limits_{b = 1}^{{{\log}_2}L} \left| {{\Lambda}_{k,t,b}}\right|}.
\end{equation}
\SetAlgoNoLine
\SetAlFnt{\small}
\SetAlCapFnt{\normalsize}
\SetAlCapNameFnt{\normalsize}
\begin{algorithm}[!tp]
\caption{SIC-Based OAMP-MMV Algorithm}\label{Algorithm_OAMP_SIC}
\begin{algorithmic}[1]
\REQUIRE ~ Noisy observation ${\bf{Y}}$, spreading code matrix ${\bf{\widetilde S}}$, the number of devices in each interference cancellation $N^{\mathrm{sic}}$, and the maximum number of SIC iterations $I_{\mathrm{max}}$.
\ENSURE Decoded data bit matrix ${\hat{\bf{X}}}^{\mathrm{sic}}$ and the set of the detected active devices ${\kappa}$.
\STATE Set $f^{\mathrm{sic}} = 1$, $i^{\mathrm{sic}} = 0$, ${\widetilde{\bf{Y}}} = {\bf{Y}}$, ${\hat{\bf{X}}}^{\mathrm{sic}} = {\bf{O}}$, and $\kappa = {\O}$.
\WHILE {$ f^{\mathrm{sic}} $}
\STATE Obtain ${\hat{\bf{X}}}$ and ${\hat{\kappa}}$ from OAMP-MMV-SSL algorithm or OAMP-MMV-ASL algorithm.
\STATE Calculate $\{{\overline{{\Lambda}}_{k}}\}_{k\in\hat{\kappa}}$ according to (\ref{Approximate LLR_mean}), and then sort them in the descent order.
\IF{${\rm{card}}\{\hat{\kappa}\}\leq N^{\mathrm{sic}}$}
\STATE $\kappa_a = \hat{\kappa}$, \ \ \ $f^{\mathrm{sic}} = 0$.
\ELSE
\STATE Choose the first $N^{\mathrm{sic}}$ largest entries of $\{{\overline{{\Lambda}}_{k}}\}_{k\in\hat{\kappa}}$, and these devices forms the set $\kappa_a$.
\ENDIF
\STATE $k \in {\kappa}_a,
{{\hat{{{\bf X}}}}_{k,:}^{\mathrm{sic}}} \stackrel{{\Xi}}{{\longleftarrow} } {{\hat{{{\bf X}}}_{k,:}}}$.
\STATE $
{{\kappa}} = {\kappa}\bigcup {\kappa_a}$.
\STATE $k \in {\kappa}_a,
\widetilde{\bf{X}}_{k,:} \stackrel{{\Xi}^{-1}}{\longleftarrow} {{\hat{{{\bf X}}}}_{k,:}^{\mathrm{sic}}}$.
\STATE $
\widetilde{\bf Y} =\widetilde{\bf Y}-\sum\limits_{k \in \kappa_a}\widetilde{\bf S}_{:,k}\widetilde{\bf X}_{k,:}$.
\IF {$i^{\mathrm{sic}} > I_{\mathrm{max}}$}
\STATE $f^{\mathrm{sic}} = 0$.
\ELSE
\STATE $i^{\mathrm{sic}} = i^{\mathrm{sic}} + 1$.
\ENDIF
\ENDWHILE
\end{algorithmic}
\end{algorithm}
The block diagram of the proposed SIC-based OAMP-MMV-SSL and OAMP-MMV-ASL algorithms is illustrated in Fig. \ref{SICB}, and the associated pseudocode is given in Algorithm \ref{Algorithm_OAMP_SIC}. More specifically, ${\hat{\bf{X}}}$ and ${\hat{\kappa}} = \{ {k|{{\hat \alpha }_k}}=1, \forall k \}$ are obtained from the OAMP-MMV-SSL or OAMP-MMV-ASL algorithm as an initial estimate. Then, in line 4, the average ALLR is calculated for activity refinement. By choosing the first $N^{\mathrm{sic}}$ largest entries of $\left\{{\overline{\Lambda}_{k}} \right\}_{k\in{{\hat{\kappa}}}}$, we obtain a subset ${\kappa}_a$ of $\hat{\kappa}$. The decoded data of the devices in ${\kappa}_a$ is considered to be more accurate compared with that of other devices in ${\hat{\kappa}}$. Next, as described in lines 10-11 of Algorithm \ref{Algorithm_OAMP_SIC}, we obtain the decoded data and indices of the devices in ${\kappa}_a$, where ${{\hat{{{\bf X}}}}^{\mathrm{sic}}}$ is the final decoded data, $\Xi$ denotes the operation of demodulation in soft decision and channel decoding. Before performing interference cancellation, the reconstructed transmitted signals ${\widetilde{{\bf{X}}}}$ is obtained in line 12, where $\Xi^{-1}$ denotes the operation of channel coding and modulation. Finally, the residual ${\widetilde{{\bf{Y}}}}$ is obtained in line 13. When the cardinality of $\kappa_a$ is smaller than $N^{\mathrm{sic}}$ or the number of SIC iterations reaches a predetermined level $I_{\mathrm{max}}$, the SIC-based OAMP-MMV-SSL algorithm or the OAMP-MMV-ASL algorithm are terminated.

\section{Performance Analysis}\label{S4}

In this section, the SE and the computational complexity of the proposed two algorithms are analyzed. The SE of the proposed OAMP-MMV-SSL algorithm and OAMP-MMV-ASL algorithm is derived for theoretically predicting the mean square error (MSE) and bit error rate (BER) performance. Also, the computational complexity of the proposed algorithms is compared against that of the state-of-the-art benchmarks.

\subsection{State Evolution}\label{S4.1}

The performance of AMP-type algorithms can be analyzed by studying the SE in the large system limit, as discussed in the previous works \cite{{AMP},{OAMP}}. Therefore, we employ the SE to theoretically predict the performance and to optimize the associated parameters for achieving the desired performance.

As far as conventional OAMP algorithms are concerned, the associated SE recursion is defined as follows
\begin{subequations} \label {SE_eq:1}
\begin{align}
 \text{LE}: { {{\tau}}^{i}} & = \dfrac{{K - M}}{M}{ {{v}}^{i-1}} + \dfrac{K}{M}{\sigma ^2}, \label{SE_eq:1A} \\
\text{NLE}:{v^i} & = {\rm{E}}\left[ {{{\left| {U - X} \right|}^2}} \right], \label{SE_eq:1B} \\
{{\vartheta^i}} & = {\rm{E}}\left[\left|{{{\rm{E}}\left[ {{X}}|{X} + \sqrt{\tau ^i}{Z} \right]}} -{X}\right|^2\right], \label{SE_eq:1C}
\end{align}
\end{subequations}
where $U = \frac{{{\tau ^i}}}{{{\tau ^i} - {\vartheta ^i}}}\left\{ {{\rm{E}}\left[ {X|X + \sqrt {{\tau ^i}} Z} \right] - \frac{{{\vartheta ^i}}}{{{\tau ^i}}}\left( {X + \sqrt {{\tau ^i}} Z} \right)} \right\}$ and ${\vartheta}$ is the predicted MSE,
${{X}} \sim p(x)$ where $p(x)$ is the same as the a priori distribution in (\ref{prior}), and $Z \sim {\cal{CN}}(z;0,1)$. According to the definition of expectation, $v^i$ and $\vartheta^i$ can be written as
\begin{subequations} \label {SE_mmse:1}
\begin{align}
& {v^i} = \displaystyle{\int} {\int {{{\rm{E}}\left[ {{{\left| {U - X} \right|}^2}} \right]{\cal D}{x}{\cal D}{z}} }}, \label{SE_mmse:1A} \\
& {{\vartheta^i}} = \displaystyle{\int} {\int {{\left|{{{\rm{E}}\left[ {{X}}|{X} + \sqrt{\tau ^i}{Z} \right]}} -{X}\right|^2{\cal D}{x}{\cal D}{z}} }}, \label{SE_mmse:1B}
\end{align}
\end{subequations}
respectively, where ${\cal D}{x} = p(x)dx$ and ${\cal D}{z} = \exp \left({-|z|^2}\right)/ \pi dz$.
\SetAlgoNoLine
\SetAlFnt{\small}
\SetAlCapFnt{\normalsize}
\SetAlCapNameFnt{\normalsize}
\begin{algorithm}[!t]
\caption{SE of OAMP-MMV Algorithms}\label{Algorithm_SE}
\begin{algorithmic}[1]
\REQUIRE ~ The number of subcarriers $M$, the number of potential devices $K$, the maximum number of iterations $I_{\mathrm{iter}}$, the number of Monte Carlo simulations $N$, and the noise variance $\sigma^2$.
\ENSURE The predicted MSE ${{\vartheta^i}}$.
\STATE ${\forall k,t,n}$: Set iteration index $i$ to 1. Initialize ${{{{v_n^0}}}} =1$. ${{\lambda^{0}_{k,t}}}$ is initialized as (\ref{sparsity_int}).
\STATE Generate Monte Carlo samples $\{{\bf{X}}_n\}^N_{n=1}$. Each ${\bf{X}}_n$ follows the true a priori distribution and the sparsity structure.
\STATE Generate standard complex Gaussian noise $\{{\bf{z}}_{t,n}\}^{T,N}_{t,n=1}$.
\FOR {$ i=1 $ to $ I_{\mathrm{iter}} $}
\STATE $\forall n$, $ {{\tau_n^{i}}}= \frac{{K - M}}{M} {{v}}_n^{i-1} + \frac{K}{M}{\sigma ^2}$.
\STATE $\forall t,n$, ${\bf{r}}_{t,n}^i = {\bf{x}}_{t.n} + \sqrt{{\tau}_n^i}{\bf{z}}_{t,n}$.
\STATE
\STATE \textbf{\%for OAMP-MMV-SSL algorithm:}
\STATE Perform lines 6-12 of Algorithm \ref{Algorithm_OAMP}.
\STATE
\STATE \textbf{\%for OAMP-MMV-ASL algorithm:}
\STATE Perform lines 6-13 of Algorithm \ref{Algorithm_OAMP_MP}.
\STATE
\STATE ${{{\vartheta^i}}} = \frac{1}{{{{KTN}}}}\sum\limits_{K = 1}^K\sum\limits_{t = 1}^T {\sum\limits_{n = 1}^{{N}} {\left| {\mu _{k,t,n}^i - x_{k,t,n}^i} \right|^2}}$.
\STATE $\forall n, {v_n^i} = \frac{1}{{{{KT}}}}\sum\limits_{t = 1}^T {\sum\limits_{k = 1}^{{K}} \left| {u_{k,t,n}^i - x_{k,t,n}^i} \right|^2}$.
\ENDFOR
\end{algorithmic}
\end{algorithm}

Limited to the SMV problem, $p(x)$ cannot characterize the sparsity structure in (\ref{sparsitypattern}), which would deteriorate the prediction accuracy of SE. To take the sparsity structure into account and to simplify the calculation of (\ref{SE_mmse:1}), we adopt Monte Carlo methods to carry out SE. Particularly, as for the SE of the OAMP-MMV-SSL algorithm, we generate enough samples of ${\bf{X}}$ according to the a priori distribution (\ref{prior}) and the sparsity structure (\ref{sparsitypattern}). Based on the assumption (\ref{NLE_base}), we add the Gaussian noise ${{\bf{Z}}}$ to ${\bf{X}}$ for obtaining ${\bf{R}}$. Then, we have the approximation
\begin{subequations} \label {SE_mmse2:1}
\begin{align}
& {v_n^i} \approx \dfrac{1}{{{{KT}}}}\sum\limits_{t = 1}^T {\sum\limits_{k = 1}^{{K}} \left| {u_{k,t,n}^i - x_{k,t,n}^i} \right|^2}, \label{SE_mmse2:1A}\\
& {{\vartheta^i}} \approx \dfrac{1}{{{{KTN}}}}\sum\limits_{n = 1}^N\sum\limits_{t = 1}^T {\sum\limits_{k = 1}^{{K}} \left| {\mu _{k,t,n}^i - x_{k,t,n}^i} \right|^2}, \label{SE_mmse2:1B}
\end{align}
\end{subequations}
where $N$ is the number of iterations. As for the OAMP-MMV-ASL algorithm, we adopt the same approach to make sure that the structured sparsity learning via message passing is taken into account. The whole SE recursions for OAMP-MMV-SSL and OAMP-MMV-ASL algorithms are summarized in Algorithm \ref{Algorithm_SE}. The accuracy of the adopted approximations will be verified in the simulations section.

\emph{Remark}: We assume the availability of the true noise variance ${\sigma^2}$ for SE prediction. Numerical results show that this assumption has a negligible impact on the capacity of the prediction performance of SE. Based on the SE recursion, the BER performance can be predicted by comparing the transmitted signal ${\bf{X}}$ and the MMSE estimate $\hat{\bf{X}}$. Numerical results demonstrate that the predicted BER performance is close to the simulated BER.

\subsection{Computational Complexity Analysis}\label{S4.2}
\begin{table*}[!tp]
\vspace{-4mm}
\renewcommand\arraystretch{1.5}
\caption{Complexity Analysis}
\centering
\begin{tabular}{c|l}
\Xhline{1pt}
Algorithm & Number of complex-valued multiplications in each iteration \\
\Xhline{1pt}
\vspace{-1mm}
OAMP-MMV-SSL & $3MKT + MT + \frac{13}{4}KTL + \frac{25}{4}KT $ \\
\vspace{-1mm}
OAMP-MMV-ASL & $3MKT + MT + \frac{13}{4}KTL + 3K{T^2} + \frac{27}{4}KT $\\
\vspace{-1mm}
GSP & $\left( {2M + 1} \right)KT + M{T^2} + 2M\left( {T + 1} \right)K_a^2 + 2K_a^3$ \\
\vspace{-1mm}
SWOMP & $MKT+M{T^2}+2MTi+2M{i^2}+i^3$ \\
\vspace{-1mm}
SAMP & $\left( {2M + 3} \right)KT + 2M\left( {T + 1} \right){j^2} + T + 2{j^3}$ \\
AMP-MMV & $4MKT + \frac{7}{4}MK + \frac{19}{2}KT $ \\
\Xhline{1pt}
\end{tabular}
\label{cc}
\vspace{1mm}
\\{Note: $i$ denotes the iteration index \cite{SWOMP} and $j$ denotes the stage index \cite{SAMP}}.
\vspace{-2mm}
\end{table*}
In Table \ref{cc}, the computational complexity of the proposed algorithms is analyzed and compared with that of state-of-the-art CS baseline algorithms. The baseline algorithms include the generalized subspace pursuit (GSP) algorithm \cite{GSP}, the simultaneous weighted orthogonal matching pursuit (SWOMP) algorithm \cite{SWOMP}, the sparsity adaptive matching pursuit (SAMP) algorithm \cite{SAMP}, and the AMP-MMV algorithm \cite{AMPG}.

We choose the number of complex-valued multiplications as the metric of interest, and the complexity of one real-valued multiplication is assumed to be equal to one quarter of the complexity of one complex-valued multiplication. From Table \ref{cc}, we evince that the GSP, SWOMP, and SAMP algorithms (which are greedy-type CS algorithms), have the same order of computational complexity, i.e., the cube of $K_a$, due to the matrix inversion needed for LS estimation. Such a high computational complexity cannot accommodate the requirements of low-delay and low processing complexity for application to future mMTC systems.

By contrast, the computational complexity of the proposed OAMP-MMV algorithms increases only linearly with $K$ and $M$. This is because the matrix inversion is avoided when $\widetilde{\bf{S}}$ is a partial DFT matrix. The computational complexity of the OAMP-MMV-ASL algorithm is higher than that of the OAMP-MMV-SSL algorithm, especially when $T$ is considerably large. Since the data packets in mMTC are usually small, the value of $T$ is usually small, and the gap in computational complexity is negligible. Simulation results illustrated next show that the higher computational complexity of the OAMP-MMV-ASL algorithm is worthy, since a considerable performance gain compared with the OAMP-MMV-SSL algorithm is obtained, especially when $T$ is small.

The two SIC-based OAMP-MMV algorithms are not compared in Table \ref{cc} due to the following reasons. During each iteration of interference cancellation, the complete execution of the OAMP-MMV-SSL or OAMP-MMV-ASL algorithms is needed for initial estimation, as shown in line 3 of Algorithm \ref{Algorithm_OAMP_SIC}. The remaining manipulations such as subtractions and demodulations give a minor contribution to the computational complexity. In view of this fact, we choose an appropriate $N^{\mathrm{sic}}$ to guarantee a small number of iterations in SIC and to reduce computational complexity. Therefore, the SIC-based OAMP-MMV algorithms have about one order of magnitude higher computational complexity than the original algorithms without SIC.

\section{Simulation Results}\label{S5}
In this section, we evaluate the performance of the proposed schemes for JADD in grant-free massive access. For the presented simulation results, we set $K=500$ potential devices in total, with $K_a=50$ of them being active in $T$ OFDM symbol intervals (i.e., one time slot). The spreading codes are chosen from a partial DFT matrix and preassigned to different devices in advance. The length of the spreading code is equal to $M$, which is also equal to the number of subcarriers for random access and to the number of measurements in the formulated CS problem. We adopt quadrature phase shift keying (QPSK) modulation for all simulations and Turbo codes{\protect\footnotemark[3]}\footnotetext[3]{The error correction capability of channel coding can reduce the error propagation of the proposed SIC-based schemes. Usually, the Turbo codes have better performance than the LDPC codes in the case of short code length and low code rate, while short-packet communication is one of the typical characteristics of mMTC\cite{TurboAdv}. Therefore, we adopt the Turbo codes in our proposed schemes.} \color{black}for channel coding in Figs. \ref{SIC}-\ref{SIC_Nsic}. As for the parameters of the SIC-based OAMP-MMV algorithms, we set the number of devices in each interference cancellation iteration $N^{\mathrm{sic}}$ multiplied by the maximum iterations $I_{\mathrm{max}}$, i.e., $N^{\mathrm{sic}}\cdot I_{\mathrm{max}}$, to $100${\protect\footnotemark[4]}. \color{black}We define the activity detection error probability (ADEP) and the BER as follows
\footnotetext[4]{Although the exact number of active devices $K_a$ is unknown, we assume that the BS has known that $K_a$ is smaller than a threshold. Therefore, we set the product of $N^{\mathrm{sic}}$ and $I_{\mathrm{max}}$, i.e., the possible maximum number of detected active devices, to $100$, which is large enough to be the threshold in our simulations.}
\begin{equation}\label{Pe_and_BER}
{{\mathrm{ADEP}}} = \dfrac{1}{K}{{\sum\limits_{k = 1}^K {\left| {{{\hat \alpha }_k} - {\alpha _k}} \right|} }},\ {\mathrm{BER}} = 1-\dfrac{{N_s}}{{K_a T {{\log }_2}L}},
\end{equation}
where $N_s$ denotes the number of correctly decoded bits for the successfully detected devices. \color{black}The number of iterations of the OAMP-MMV algorithms $I_{\mathrm{iter}}$ is set to 50. Moreover, to compare the simulation results with SE-based results, we adopt the MSE metric, which is defined as
\begin{equation}\label{MSE}
{\mathrm{MSE}} = \dfrac{1}{{KT}}\sum\limits_{t = 1}^T{\rm{E}}\left[ { \left\| {{\bf{\hat x}}_t - {\bf{x}}_t} \right\|_2^2} \right],
\end{equation}
and set $N = 1000$. The baseline algorithms include the GSP algorithm, SWOMP algorithm, SAMP algorithm, Oracle LS algorithm \cite{TWC_DY}, AMP-MMV algorithm, and Gene-Aided OAMP algorithm,
\begin{figure}
\centering
\vspace*{-2mm}
\hspace*{-2mm}
{\includegraphics[width=8 cm]
{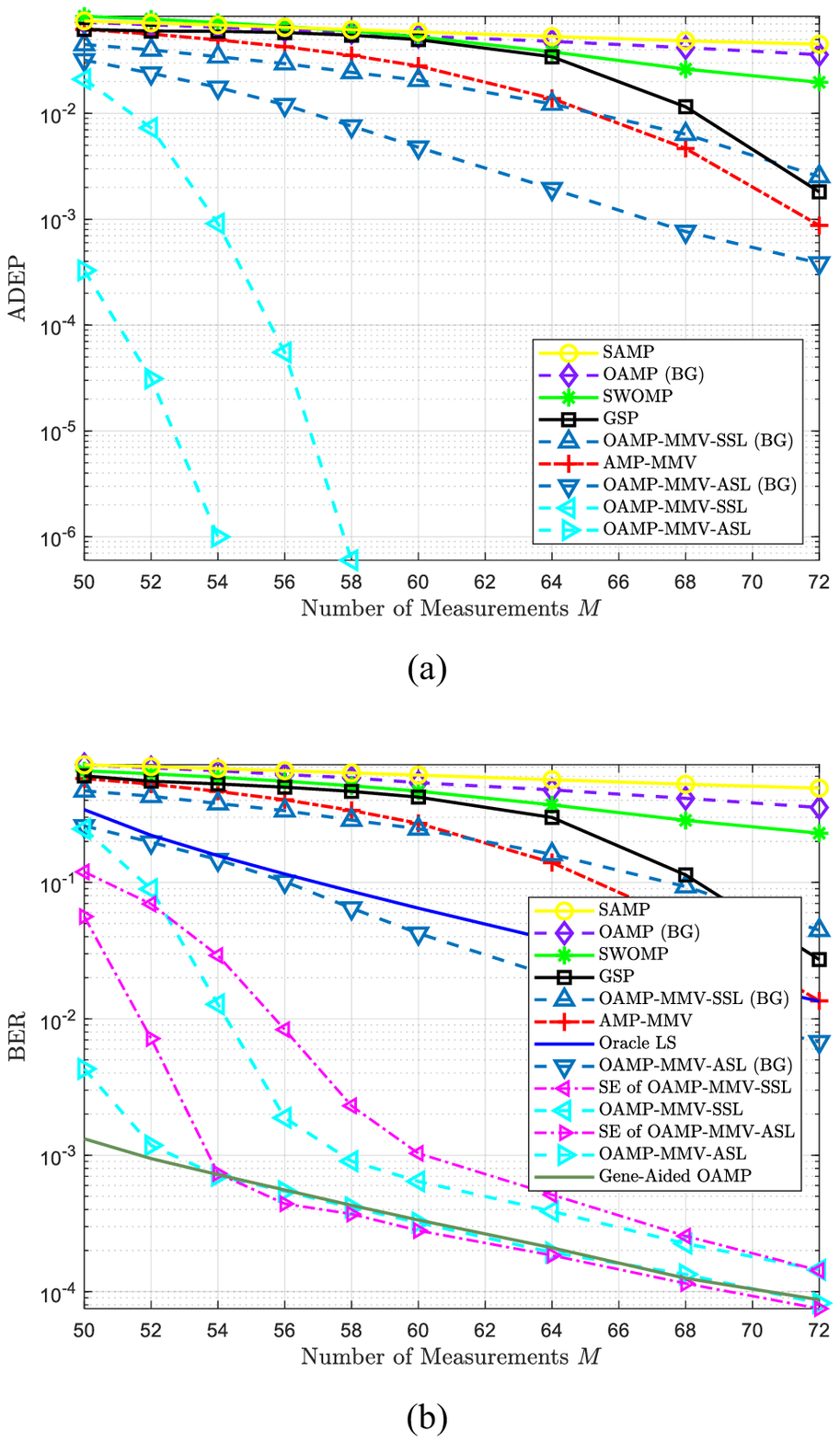}}
\captionsetup{font={footnotesize}, singlelinecheck = off, justification = raggedright,name={Fig.},labelsep=period}
\centering
\caption{Comparison of different detection algorithms{\protect\footnotemark[5]} versus $M$ given $T = 10$ and SNR $=10$ dB: (a) ADEP; (b) BER.}
\label{Simulation_M}
\vspace*{-4mm}
\end{figure}
{\footnotetext[5]{\color{black}In Fig. \ref{Simulation_M}, we substitute $v^i_t = \frac{\|{\bf y}_t-{\bf F}{\bf u}^i_t\|_2^2-M\sigma^2}{M}$ of [27, (30)] into the variance ${\tau}_t^{i} = \frac{K-M}{M}{v}_t^{i-1} + \frac{K}{M}({\sigma^2})^{i-1}$ in line 4 of Algorithm 2 for better detection performance of the proposed OAMP-MMV-ASL algorithm (mainly for the case of low number of observations, i.e., $M<56$, which leads to the gap between the SE and the simulation results).}}where the AMP-MMV algorithm applies (\ref{sparsity_MMV}) to leverage the sparsity structure. The Gene-Aided OAMP algorithm is the proposed OAMP-MMV algorithm knowing perfect support and true noise variance. Specifically, in each iteration of the Gene-Aided OAMP algorithm, we have $\lambda_{k,t}=\pi_{k,t}=1$ for active devices and $\lambda_{k,t}=\pi_{k,t}=0$ for inactive devices.
Therefore, it is the low-bound of our proposed OAMP-MMV-SSL/ASL algorithms. The Matlab codes of this paper will be released on the corresponding author's homepage {\em https://gaozhen16.github.io/\#Latest} in the near future.

\color{black}
\begin{figure}[!tp]
\centering
\vspace*{-4mm}
\hspace*{-2mm}
\includegraphics[width=8 cm]
{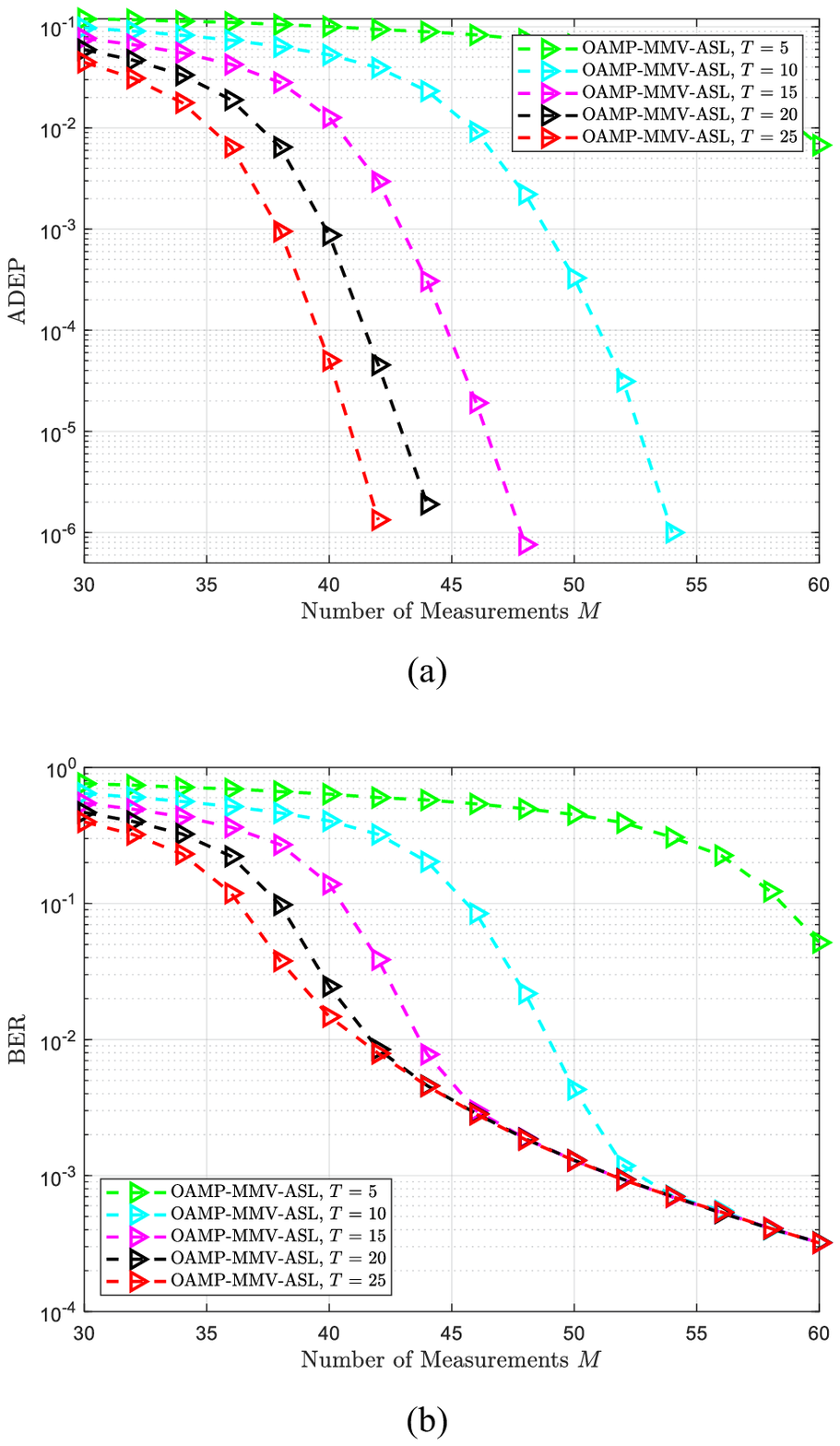}
\centering
\captionsetup{font={footnotesize}, singlelinecheck = off, justification = raggedright,name={Fig.},labelsep=period}
\caption{Comparison of the OAMP-MMV-ASL algorithm versus $M$ under different $T$, given SNR $= 10$ dB: (a) ADEP; (b) BER.}
\label{Overloading}
\vspace*{-3mm}
\end{figure}
Fig. \ref{Simulation_M} compares the ADEP and BER performance of different schemes versus the spreading code length $M$, where the conventional OAMP algorithm and the proposed OAMP-MMV algorithm that employs a Bernoulli-Gaussian (BG) a priori distribution are considered. Compared with other algorithms, the OAMP-MMV-SSL and OAMP-MMV-ASL algorithms provide better performance.
By contrast, the performance of the two proposed OAMP-MMV algorithms based on the BG a priori distribution is worse, since the prior knowledge of discrete modulation symbols is not fully exploited. Even though, it can seen that they outperform the conventional OAMP algorithm, which owes to the utilization of sparsity structure. We observe a
performance gap between the OAMP-MMV-SSL and OAMP-MMV-ASL algorithms, especially when $M$ is small. This is because the latter algorithm can learn the structured sparsity more accurately, at the cost of a higher computational complexity. Besides, it can be observed that when $M$ is larger than $52$, the BER performance of the proposed OAMP-MMV-ASL algorithm is almost overlapped with that of the Gene-Aided OAMP algorithm. \color{black} When the number of observation is large enough, i.e., $M>58$, the two proposed OAMP-MMV algorithms can detect the active device with a high probability of success, and their BER gap becomes small. The SE results of the two proposed OAMP-MMV algorithms are presented in Fig. \ref{Simulation_M}(b),
which shows that the BER performance is well predicted when $M$ is larger than $60$. Moreover, the baseline algorithms need more observations to obtain good performance. The Oracle LS that assumes the perfect activity detection is the low-bound of the greedy algorithms above. However, it is still inferior to the proposed OAMP-MMV-SSL/ASL algorithms in terms of BER performance, since it cannot utilize the prior information of signals' discrete constellation modulation in the stage of data detection. \color{black}

\begin{figure}[!tp]
\centering
\vspace*{-4mm}
\hspace*{-2mm}
\includegraphics[width=8 cm]
{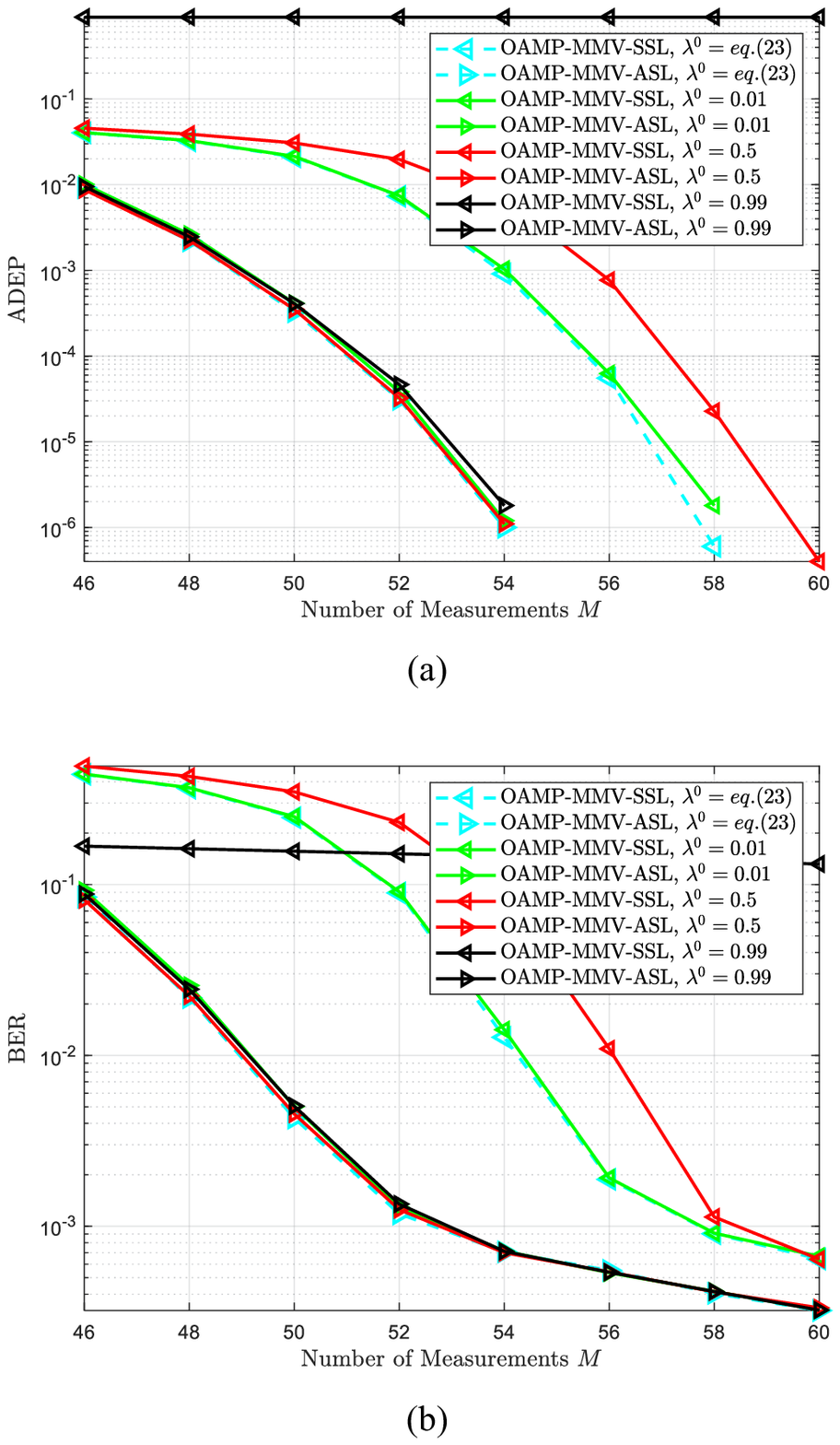}
\centering
\captionsetup{font={footnotesize}, singlelinecheck = off, justification = raggedright,name={Fig.},labelsep=period}
\caption{Given SNR $= 10$ dB and $T = 10$, the detection performance of the OAMP-MMV-SSL/ASL algorithms versus $M$ with different initializations of the sparsity ratio $\lambda^0$: (a) ADEP; (b) BER.}
\label{EM_lambda}
\vspace*{-3mm}
\end{figure}

Given $\text{SNR}=10$ dB, Fig. 6 compares the performance of the proposed OAMP-MMV-ASL algorithm under a small number of measurements with different $T$. It is worth noting that the OAMP-MMV-ASL algorithm can obtain good detection performance even if it is overloading (i.e., $M<K_a$), as long as $T$ is large enough. For example, when $T=25$, the proposed scheme can achieve a good ADEP of $10^{-4}$ and BER of $10^{-2}$ under the overloading rate $K_a/M = 125\%$. This result can be further improved by adding channel coding and SIC processing into the proposed scheme. Besides, it can be seen that with the increase of $T$, the OAMP-MMV-ASL algorithm arriving at a high probability of successful detection has a lower requirement on $M$. Particularly, if we define the high probability of success is ADEP $> 10^{-6}$, the required minimum number of measurements $M_{\mathrm{min}}$ is $43$, $45$, and $48$ when $T$ is equal to $25$, $20$, and $15$, respectively, given SNR $=10$ dB.

\begin{figure}[!tp]
\centering
\vspace*{-4mm}
\hspace*{-2mm}
\includegraphics[width=8 cm]
{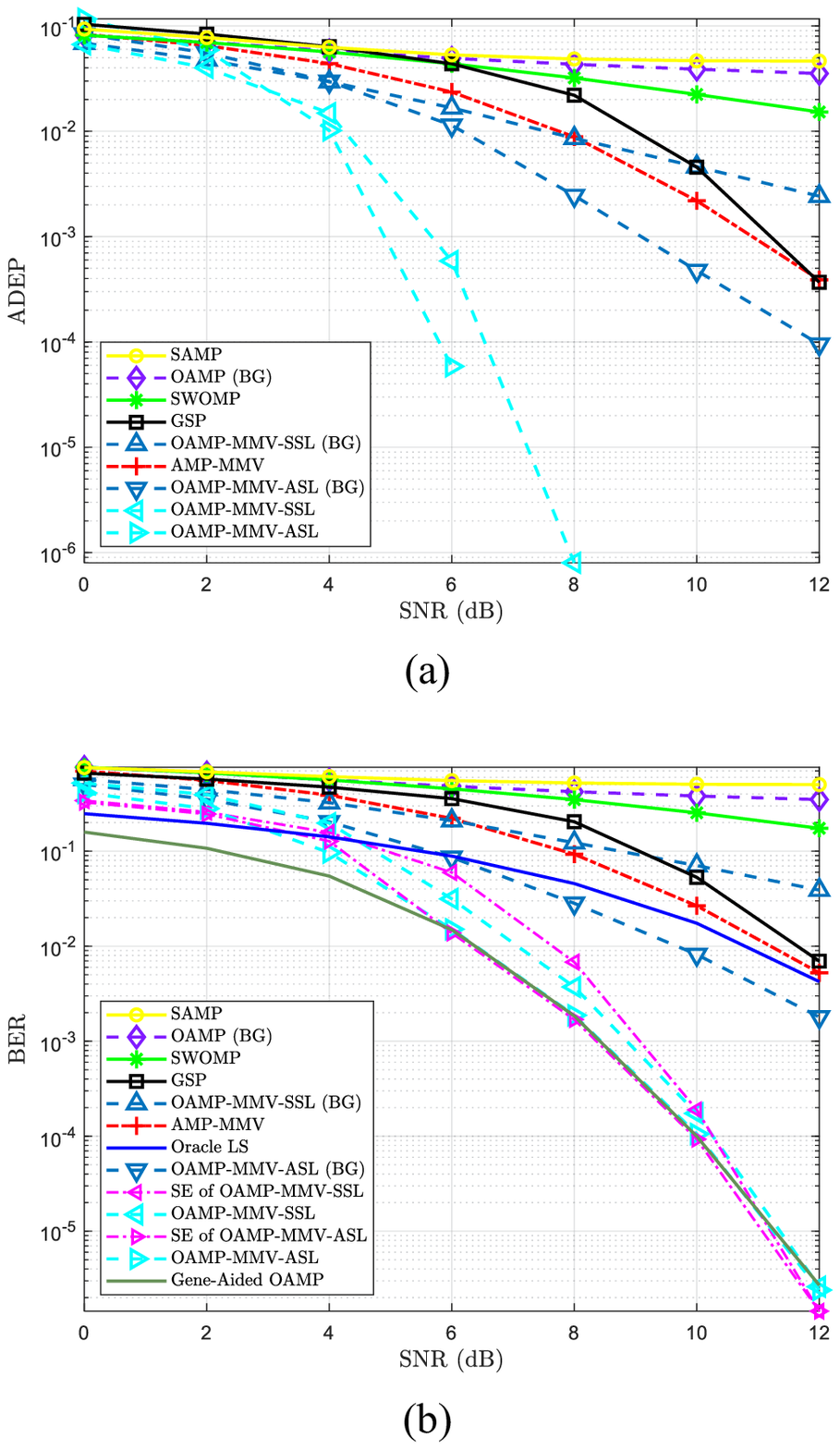}
\centering
\captionsetup{font={footnotesize}, singlelinecheck = off, justification = raggedright,name={Fig.},labelsep=period}
\caption{Comparison of different detection algorithms versus SNR given $T = 10$ and $M =70$: (a) ADEP; (b) BER.}
\label{Simulation_SNR}
\vspace*{-3mm}
\end{figure}
The performance comparison of the proposed schemes under different initializations of EM parameters is provided in Fig. \ref{EM_lambda}, given SNR $=10$ dB and $T=10$. Both the proposed algorithms are robust to the initialization of noise variance $({\sigma^2})^0$, so we only provide the numerical results with different ${\lambda}^0$. For interesting readers, we recommend the complete comparison on the arXiv website (https://arxiv.org/abs/2011.07928). In Fig. \ref{EM_lambda}, $({\sigma^2})^0$ is initialized as equation (24), and ${\lambda}^0$ is set to $0.01$, $0.5$, $0.99$, and according to equation (23), respectively. Among these four settings, we can see that the last one has the best detection performance for the OAMP-MMV-SSL algorithm. ${\lambda}^0 = 0.5$ represents that there is no prior information about the sparsity, leading to the performance deterioration of the OAMP-MMV-SSL algorithm. When ${\lambda}^0$ increases to $0.99$, it does not work at all. On the contrary, the OAMP-MMV-ASL algorithm works well, no matter how ${\lambda}^0$ changes. These results indicate that for the OAMP-MMV-SSL algorithm, the EM method can perform well in estimating the sparsity ratio unless its initialization is extremely deviated. The OAMP-MMV-ASL algorithm is more robust to the change of ${\lambda}^0$, since it has a stronger capability of learning the sparsity structure than the OAMP-MMV-SSL method.

\color{black}
\begin{figure}[!tp]
\centering
\vspace*{-4mm}
\hspace*{-2mm}
\includegraphics[width=8 cm]
{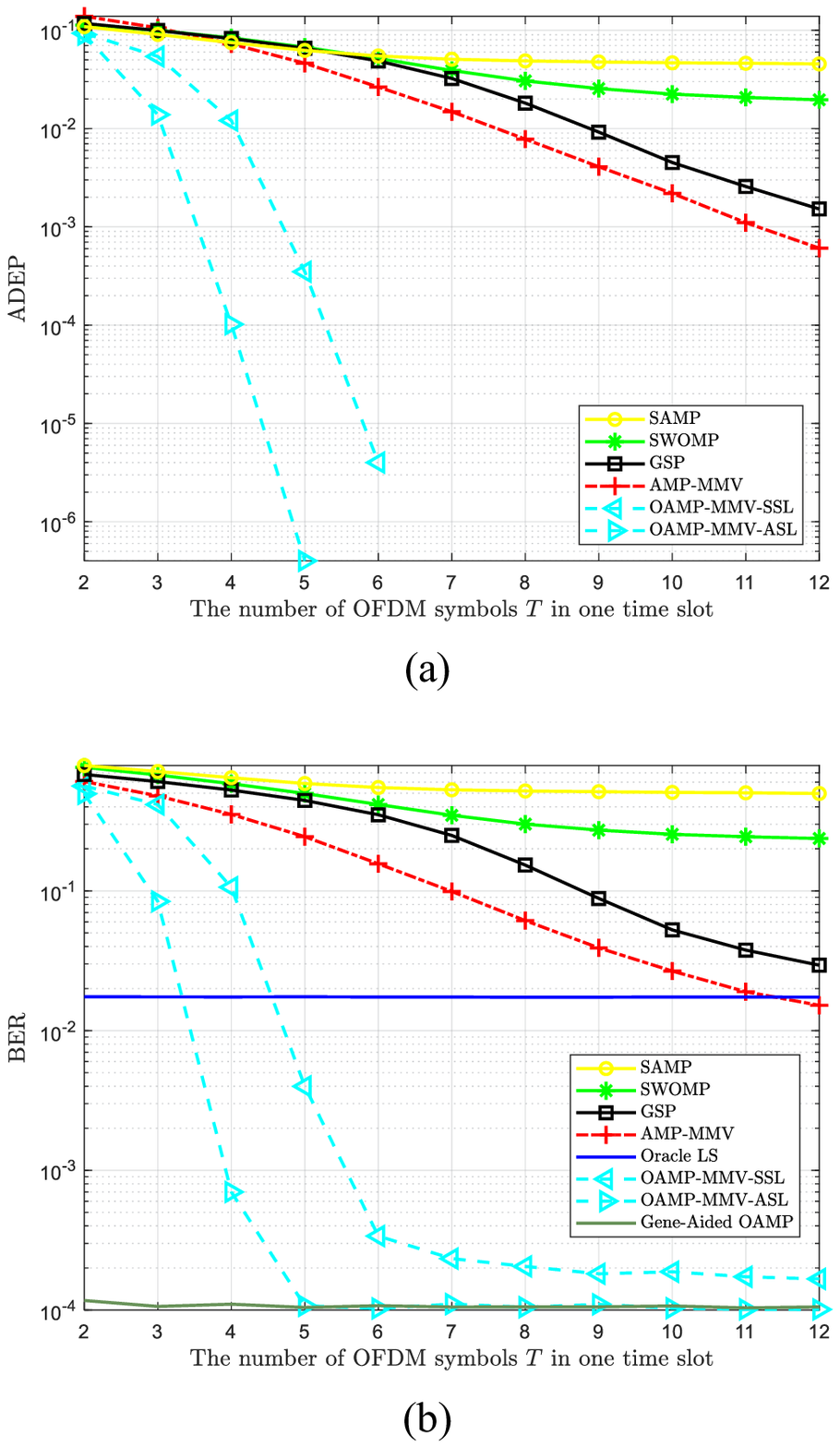}
\centering
\captionsetup{font={footnotesize}, singlelinecheck = off, justification = raggedright,name={Fig.},labelsep=period}
\caption{Comparison of different detection algorithms versus $T$ given $M = 70$ and SNR $= 10$ dB: (a) ADEP; (b) BER.}
\label{Simulation_T}
\vspace*{-3mm}
\end{figure}
Fig. \ref{Simulation_SNR} compares the ADEP and BER performance of different schemes as a function of the SNR. It can be observed that, when the SNR is large, both ADEP and BER of our proposed OAMP-MMV algorithms decrease rapidly and outperform other algorithms significantly. The performance of the OAMP-MMV-SSL algorithm is close to that of the OAMP-MMV-ASL algorithm when the SNR is larger than $8$ dB, since the sparsity structure in the high SNR range can be well learned. In addition, Fig. \ref{Simulation_SNR}(b) demonstrates that the simulated BER performance can be predicted by the SE results in the high SNR regime. It is worth noting that the proposed OAMP-MMV-ASL algorithm has the same BER performance as that of the Gene-Aided OAMP algorithm when SNR $>6$ dB. \color{black}However, the greedy algorithms have poor performance even when the SNR is relatively high, as they fail to fully exploit the prior knowledge of transmitted signals and the sparsity structure.

\begin{figure}[!tp]
\centering
\vspace*{-4mm}
\hspace*{-2mm}
\includegraphics[width=8 cm]
{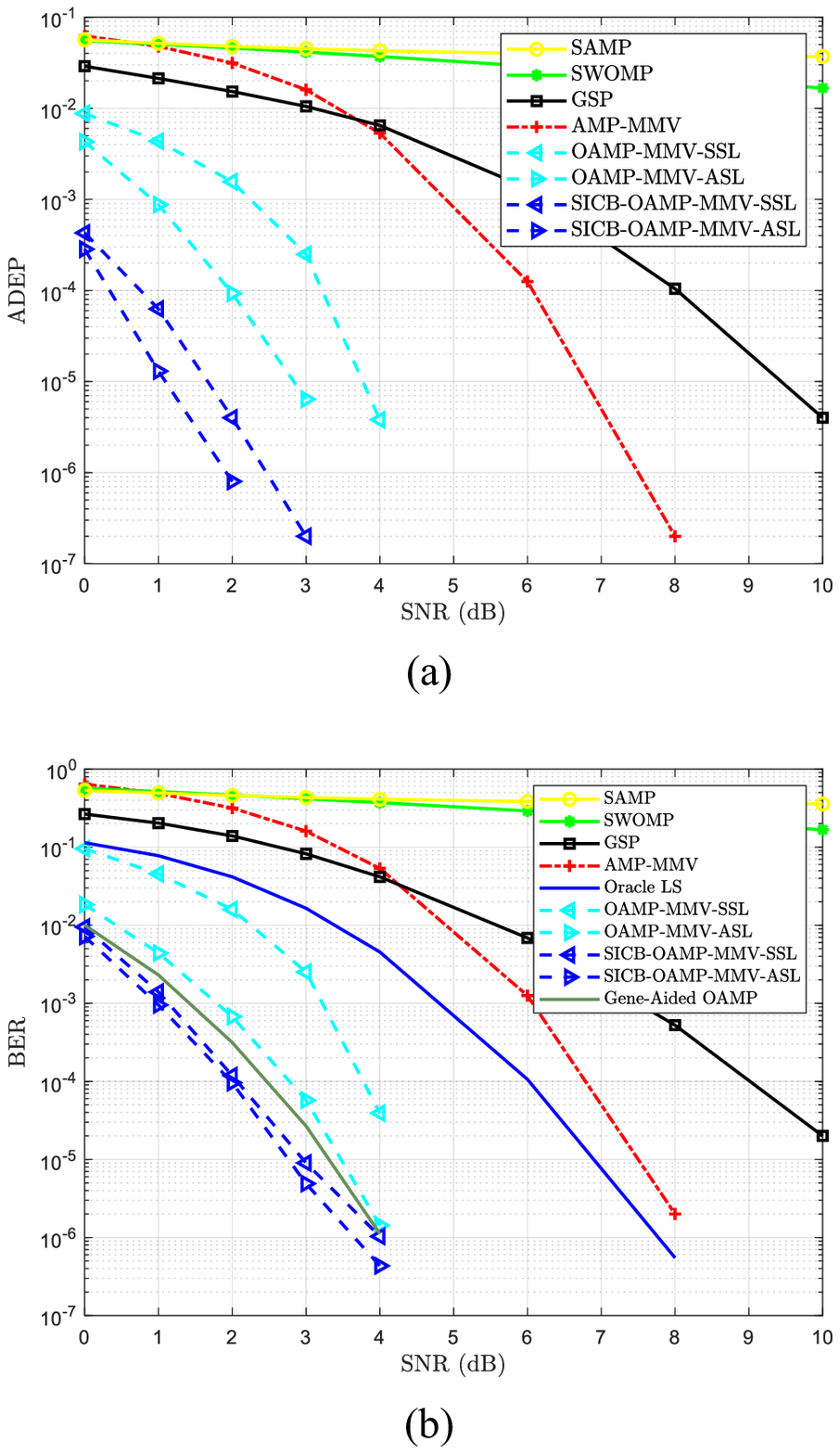}
\centering
\captionsetup{font={footnotesize}, singlelinecheck = off, justification = raggedright,name={Fig.},labelsep=period}
\caption{Gains of the proposed SIC-based OAMP-MMV algorithms versus SNR given $M = 70$, $T = 51$: (a) ADEP; (b) BER.}
\label{SIC}
\vspace*{-3mm}
\end{figure}
Fig. \ref{Simulation_T} compares the ADEP and BER performance of the considered schemes as a function of the number $T$ of OFDM symbols in each time slot. All algorithms perform poorly when $T = 2$. This is because the MMV model boils down to the SMV model when $T$ is close to $1$, resulting in limited gains when exploiting the sparsity structured in (\ref{sparsitypattern}). As $T$ increases, the OAMP-MMV-ASL and OAMP-MMV-SSL algorithms distinctively outperform other baseline algorithms. The data detection performance of the Oracle LS does not improve with the increase of $T$. Though it knows all active devices, its BER performance is about two orders of magnitude worse than our proposed schemes when $T>8$. \color{black}In addition, Fig. \ref{Simulation_T} demonstrates that the OAMP-MMV-ASL algorithm can learn the sparsity structure more accurately than the OAMP-MMV-SSL algorithm, especially when $T$ is small. In particular, when $T$ is larger than 7, the two proposed OAMP-MMV algorithms can achieve almost perfect activity identification, as shown in Fig. \ref{Simulation_T}(a), which verifies the significant gain benefited from exploiting the sparsity structure. This observation indicates that it is better to adopt the OAMP-MMV-SSL algorithm when $T$ is large, which has lower computational complexity and negligible BER performance loss compared to the OAMP-MMV-ASL algorithm. In contrast, the OAMP-MMV-ASL algorithm is more suitable if $T$ is small. A BER floor of the proposed algorithms can be observed in Fig. \ref{Simulation_T}(b). This is because the device's activity can be almost perfectly obtained if $T>7$, while the fixed SNR is the key limiting factor for BER improvement.

\begin{figure}[!tp]
\centering
\vspace*{-4mm}
\hspace*{-2mm}
\includegraphics[width=8 cm]
{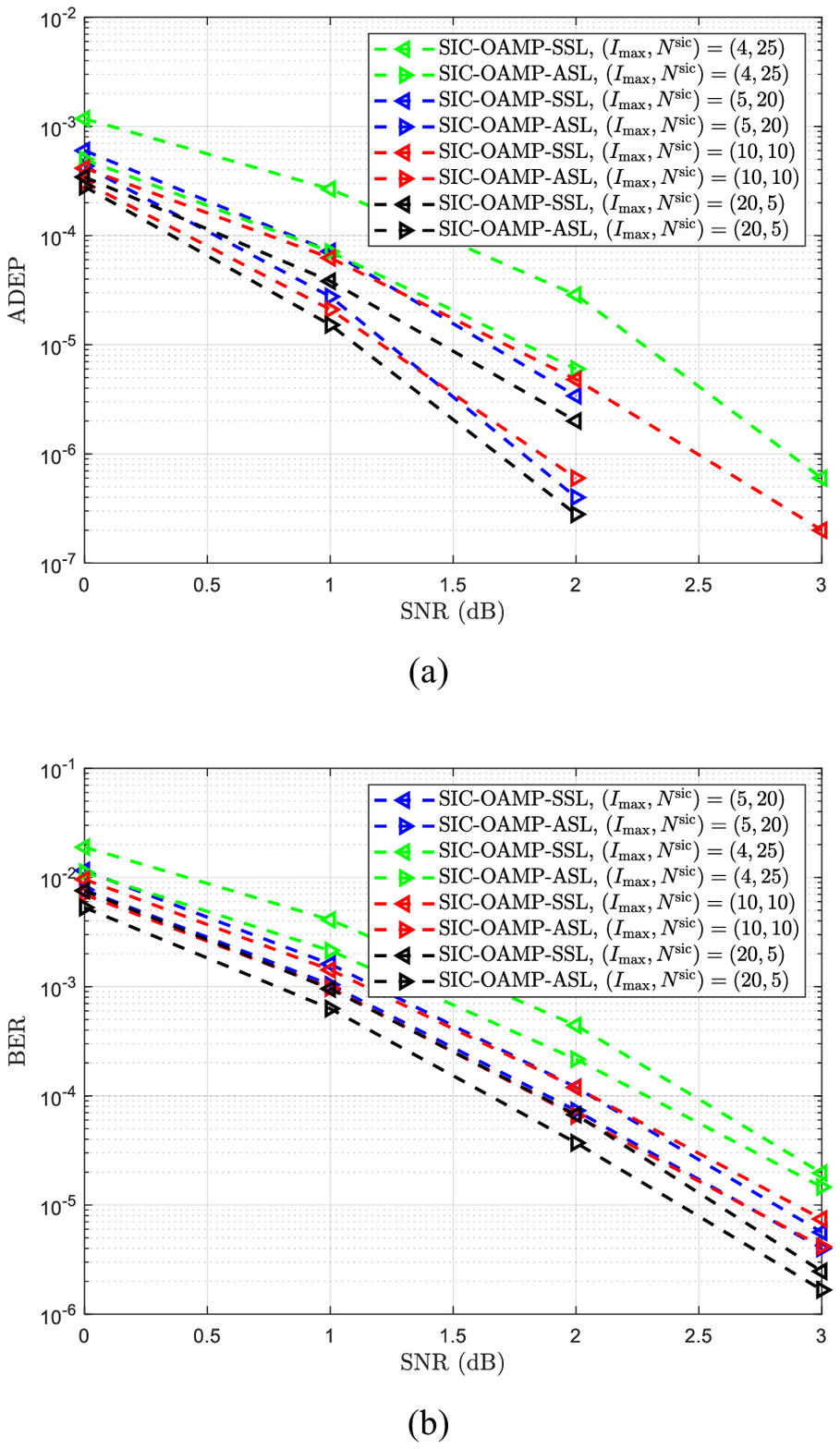}
\centering
\captionsetup{font={footnotesize}, singlelinecheck = off, justification = raggedright,name={Fig.},labelsep=period}
\caption{The comparison of detection performance of the SIC-based OAMP-MMV algorithms versus SNR with different pairs of $N^{\mathrm{sic}}$ and $I_{\mathrm{max}}$, given $M = 70$ and $T = 51$: (a) ADEP; (b) BER.}
\label{SIC_Nsic}
\vspace*{-3mm}
\end{figure}
Fig. \ref{SIC} shows the effectiveness of the two proposed SIC-based OAMP-MMV algorithms with $N^{\mathrm{sic}}=I_{\mathrm{max}}=10$, where all schemes consider Turbo coding. In particular, a code rate equal to $1/3$, a code length of $30$, and a number of tail bits of $12$ are considered. It can be observed that the proposed SIC-based schemes achieve better ADEP and BER performance than the OAMP-MMV-SSL and OAMP-MMV-ASL algorithms. The two proposed SIC-based schemes have similar BER performance, even though their ADEP gap is still apparent. Therefore, we conclude that the SIC-based OAMP-MMV-SSL algorithm is more practical considering the offered trade-off in terms of performance and computational complexity.

In order to investigate the performance difference of the proposed SIC-based schemes under different $N^{\mathrm{sic}}$, we provide the ADEP and BER performance in Fig. 11, where $N^{\mathrm{sic}}\cdot I_{\mathrm{max}}=100$ is considered. It can be observed that the detection performance of the proposed SIC-based OAMP-MMV-SSL/ASL algorithms under $(I_{\mathrm{max}}, N^{\mathrm{sic}}) = (20, 5)$ is superior to that under other conditions, and the worst performance is obtained under the case of $I_{\mathrm{max}} = 4$ and $ N^{\mathrm{sic}} =  25$. This indicates that the performance of the proposed SIC-based schemes becomes better when $N^{\mathrm{sic}}$ decreases. However, the selection of $N^{\mathrm{sic}}$ does not have an obvious impact on the detection performance when $10\leq N^{\mathrm{sic}}\leq 20$.

\begin{figure}[!tp]
\centering
\vspace*{-4mm}
\hspace*{-2mm}
\includegraphics[width=8 cm]
{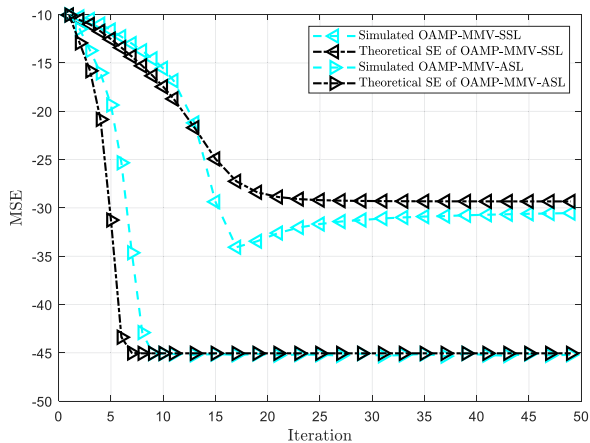}
\centering
\captionsetup{font={footnotesize}, singlelinecheck = off, justification = raggedright,name={Fig.},labelsep=period}
\caption{MSE comparison between the simulated results and theoretical SE, given $M=70$, SNR $= 10$ dB, and $T =10$.}
\label{SE}
\vspace{-3mm}
\end{figure}

\color{black}
The comparison of the simulated results and theoretical SE is shown in Fig. \ref{SE}. It can be observed that the simulated results can be well predicted by the SE results. Besides, the proposed two OAMP-MMV algorithms converge very fast, especially for the OAMP-MMV-ASL algorithm (in less than $10$ iterations). The MSE gap between the OAMP-MMV-SSL algorithm and OAMP-MMV-ASL algorithm demonstrates the better detection performance of the latter algorithm.

\section{Conclusions}\label{S6}
In this paper, a slotted grant-free massive random access scheme was proposed for enabling mMTC in future massive IoT, where the JADD problem was investigated. Specifically, we assumed that the devices' activity remain unchanged in $T$ successive OFDM symbols (i.e., one time slot), and the uplink massive access at the BS can be formulated as an MMV CS problem by utilizing the inherently structured sparsity. Different from the conventional OAMP algorithm limited to the SMV problem, the proposed OAMP-MMV-SSL and OAMP-MMV-ASL algorithms can capture the structured sparsity for improved JADD performance. Particularly, the former enjoys lower computational complexity, while the latter has a better activity detection performance, especially when $T$ is small. Considering the practical application of our proposed algorithms, the OAMP-MMV-ASL algorithm can be applied when $T$ is small, while the OAMP-MMV-SSL algorithm is better if $T$ is large. To further improve the performance, SIC-based methods were integrated into the two proposed algorithms and channel coding was taken into account. In particular, the LLR obtained in soft-decision was applied to refine the estimation of the active devices. Besides, the SE was derived to predict the performance of our algorithms. Simulation results confirmed the superiority of the proposed algorithms over state-of-the-art solutions, and demonstrated the gain of the proposed SIC-based JADD scheme, especially for the SIC-based OAMP-MMV-SSL algorithm.

\appendix
\subsection{Derivation of EM Update}\label{AP.1}
Firstly, we define an intermediate variable ${\bf{E}} = {\bf{\widetilde S}}{{\bf{X}}}$ for simplicity. Moreover, $e_{m,t}$ denotes the $(m,t)$-th element of matrix $\bf{E}$, and $\widetilde{s}_{m,k}$ denotes the $(m,k)$-th element of matrix $\widetilde{\bf{S}}$. In OAMP-MMV-SSL algorithm, we have the following relation
\begin{align}\label{EM_ln_joint}
\ln p\left( {{\bf{X}},{\bf{Y}}} \right) &= \ln p\left( {{\bf{Y}}|{\bf{X}}} \right) + \ln p\left( {\bf{X}} \right) \nonumber \\
&= \prod\limits_{m = 1}^M {\prod\limits_{t = 1}^T} {\cal CN}\left( {{y_{m,t}};{{e}_{{m,t}}},{\sigma ^2}} \right) + \prod\limits_{k = 1}^K {\prod\limits_{t = 1}^T {p\left( {{x_{k,t}}} \right)} }.
\end{align}
The posterior distribution $p\left( {{\bf{X}}|{\bf{Y}}};{{\bm{{\theta }}} ^{i-1}} \right)$ can be approximated as
\begin{equation}\label{EM_post}
p\left( {{\bf{X}}|{\bf{Y}}};{{\bm{{\theta }}} ^{i-1}} \right) \approx \prod\limits_{k = 1}^K {\prod\limits_{t = 1}^T {p\left( {{x_{k,t}}|{r^i_{k,t}}} \right)} },
\end{equation}
where ${p( {{x_{k,t}}|{r^i_{k,t}}})}$ is defined in (\ref{posterior}).

Then, we derive the update of noise variance $\sigma^2$ in OAMP-MMV-SSL algorithm. Substituting (\ref{EM_ln_joint}) into (\ref{E_step}), we have
\begin{align}\label{EM_Q_noise}
\vspace{-1mm}
Q\left( {{\bm{{\theta }}}, {{\bm{{\theta }}} ^{i-1}}} \right) = &- MT\ln {\sigma ^2} \nonumber \\
& \ + \sum\limits_{m = 1}^M {\sum\limits_{t = 1}^T {\frac{{{\rm{E}}\left[ {{{\left| {{y_{m,t}} - {{e}_{{m,t}}}} \right|}^2}|{\bf{Y}};{{\bm{{\theta }}} ^{i-1}}} \right]}}{{{\sigma ^2}}}} },
\end{align}
where the irrelevant term with respect to $\sigma^2$ has been ignored. By setting the derivative of (\ref{EM_Q_noise}) with respect to $\sigma^2$ to zero, we obtain the update rule in (\ref{noisevar_up}) as
\begin{align}\label{EM_noise}
{\left( {{\sigma ^2}} \right)^{i}}
& = \dfrac{1}{{MT}}\sum\limits_{m = 1}^M {\sum\limits_{t = 1}^T {{\rm{E}}\left[ {{{\left| {{y_{m,t}} - {e_{m,t}}} \right|}^2}|{\bf{Y}};{{\bm{{\theta }}} ^{i-1}}} \right]} } \notag \\
& = \dfrac{1}{{MT}}\sum\limits_{m = 1}^M \sum\limits_{t = 1}^T {\Big\{ {{{\left| {{y_{m,t}} - {\rm{E}}\left[ {{e_{m,t}}|{\bf{Y}};{{\bm{{\theta }}} ^{i-1}}} \right]} \right|}^2}}}  \notag \\
& \ \ \ + {{{\mathop{\rm var}} \left[ {{e_{m,t}}|{\bf{Y}};{{\bm{{\theta }}} ^{i-1}}} \right]} \Big\}},
\end{align}
where $p\left( {{\bf{X}}|{\bf{Y}}};{{\bm{{\theta }}} ^{i-1}} \right)$ can be approximated by (\ref{EM_post}), and
\begin{align}
{\rm{E}}\left[ {{e_{m,t}}|{\bf{Y}};{{\bm{{\theta }}} ^{i-1}}} \right] & = {\rm{E}}\left[ {\sum\limits_{k = 1}^K {{\widetilde{s}_{m,k}}{x_{k,t}}} |{\bf{Y}};{{\bm{{\theta }}} ^{i-1}}} \right] = \sum\limits_{k = 1}^K {{\widetilde{s}_{m,k}}\mu _{k,t}^i}, \label{EM_e_postmean} \\
{\mathop{\rm var}} \left[ {{e_{m,t}}|{\bf{Y}};{{\bm{{\theta }}} ^{i-1}}} \right] & = \sum\limits_{k = 1}^K {{{\left| {\widetilde{s}_{m,k}} \right|}^2}{\mathop{\rm var}} \left[ {{x_{k,t}}|{\bf{Y}};{{\bm{{\theta }}} ^{i-1}}} \right]} = \dfrac{1}{K}\sum\limits_{k = 1}^K {\gamma _{k,t}^i}. \label{EM_e_postvar}
\end{align}
The calculation of (\ref{EM_e_postmean})-(\ref{EM_e_postvar}) exploits the definition of the auxiliary matrix $\bf{E}$ and partial DFT matrix $\widetilde{\bf{S}}$.

\newcounter{mytempeqncnt}
\begin{figure*}[!t]
\normalsize
\setcounter{mytempeqncnt}{\value{equation}}
\setcounter{equation}{52}
\begin{equation}\label{EM_Q_lambda2}
\vspace{-2.5mm}
\dfrac{{\partial \ln p\left( {{\bf{X}},{\bf{Y}}} \right)}}{{\partial {\lambda _{k,t}}}} = \frac{\partial }{{\partial {\lambda _{k,t}}}}\left[ {\sum\limits_{k = 1}^K {\sum\limits_{t = 1}^T {\ln p\left( {{x_{k,t}}} \right)} } } \right] = \mathop {\lim }\limits_{\epsilon \to 0^+} \underbrace {\frac{{ - {\cal CN}\left( {{x_{k,t}};0,\epsilon } \right) + \left( {1/L} \right)\sum\limits_{l = 1}^L {{\cal CN}\left( {{x_{k,t}};{a_l},\epsilon } \right)} }}{{\left( {1 - {\lambda _{k,t}}} \right){\cal CN}\left( {{x_{k,t}};0,\epsilon } \right) + \left( {{\lambda _{k,t}}/L} \right)\sum\limits_{l = 1}^L {{\cal CN}\left( {{x_{k,t}};{a_l},\epsilon } \right)} }} }_{d\left( {{x_{k,t}}} \right)}.
\end{equation}
\end{figure*}
The derivation of $\lambda_{k,t}$ in (\ref{sparsity_MMV}) is described as follows. Since the Dirac function is discontinuous, we adopt the following approximation: ${\cal CN}(x;x_0,\epsilon) \rightarrow \delta(x-x_0)$, when $\epsilon\rightarrow0^+$. Under this assumption, by exchanging the order of the integral and the derivative, we have
\begin{equation}\label{EM_Q_lambda}
\dfrac{{\partial Q\left( {{\bm{{\theta }}}, {{\bm{{\theta }}} ^{i-1}}} \right)}}{{\partial {\lambda _{k,t}}}} = {\rm{E}}\left[ {\left. {\dfrac{{\partial \ln p\left( {{\bf{X}},{\bf{Y}}} \right)}}{{\partial {\lambda _{k,t}}}}} \right|{\bf{Y}};{{\bm{{\theta }}} ^{i-1}}} \right],
\end{equation}
where the derivative can be simplified as equation (\ref{EM_Q_lambda2}).
Based on (\ref{EM_Q_lambda2}), we obtain
\vspace{-1mm}
\begin{align}
\mathop {\lim }\limits_{\epsilon \to 0^+} d\left( 0 \right) & = \mathop {\lim }\limits_{\epsilon \to 0^+} \dfrac{{ - {\cal CN}\left( {0;0,\epsilon } \right)}}{{\left( {1 - {\lambda _{k,t}}} \right){\cal CN}\left( {0;0,\epsilon } \right)}} = \dfrac{{ - 1}}{{1 - {\lambda _{k,t}}}}, \label{EM_d_0} \\
\mathop {\lim }\limits_{\epsilon \to 0^+} d\left( {{a_l}} \right) & = \mathop {\lim }\limits_{\epsilon \to 0^+} \dfrac{{{\cal CN}\left( {{a_l};{a_l},\epsilon } \right)}}{{{\lambda _{k,t}}{\cal CN}\left( {{a_l};{a_l},\epsilon } \right)}} = \dfrac{1}{{{\lambda _{k,t}}}}, \ \ \ \forall l. \label{EM_d_al}
\end{align}
Substituting (\ref{EM_post}) and (\ref{EM_Q_lambda2})-(\ref{EM_d_al}) into (\ref{EM_Q_lambda}), we have
\begin{equation}\label{EM_Q_lambda3}
\vspace{-1mm}
\dfrac{{\partial Q\left( {{\bm{{\theta }}}, {{\bm{{\theta }}} ^{i-1}}} \right)}}{{\partial {\lambda _{k,t}}}}
= - \dfrac{{1 - \pi _{k,t}^i}}{{1 - {\lambda _{k,t}}}} + \dfrac{{\pi _{k,t}^i}}{{{\lambda _{k,t}}}}.
\end{equation}
By setting (\ref{EM_Q_lambda3}) equal to zero, we obtain the update rule in (\ref{sparsity_up}).

Following similar steps, the update rule of OAMP-MMV-ASL algorithm can also be derived, which is omitted for brevity. However, the update of $\lambda_{k}$, given as follows
\begin{equation}\label{EM_ASL_lambda}
\lambda_{k}^i = \frac{{\lambda _k^{i-1}\prod\limits_{t=1}^T {\eta _{k,t}^i} }}{{\left( {1 - \lambda _k^{i-1}} \right)\prod\limits_{t=1}^T {\left( {1 - \eta _{k,t}^i} \right)} + \lambda _k^{i-1}\prod\limits_{t=1}^T {\eta _{k,t}^i} }},
\end{equation}
heavily depends on the accuracy of each $\eta_{k,t}$, where $\eta_{k,t}$ is given in (\ref{yeta}). When some $\eta_{k,t}$ is wrong due to the noise, the $\lambda_{k}^i$ calculated by equation (\ref{EM_ASL_lambda}) can be opposite to the true activity pattern, leading to the deterioration of the detection performance. Therefore, we adopt the equation below
\begin{align}\label{ASL_lambda}
\lambda_{k}^i &= \frac{1}{T}\sum\limits_{t=1}^T{\xi^i_{k,t}} \notag \\
&= \frac{1}{T}\sum\limits_{t=1}^T \frac{{\lambda _k^{i-1}\prod\limits_{t' \in \backslash t} {\eta _{k,t'}^i} }}{{\left( {1 - \lambda _k^{i-1}} \right)\prod\limits_{t' \in \backslash t} {\left( {1 - \eta _{k,t'}^i} \right)} + \lambda _k^{i-1}\prod\limits_{t' \in \backslash t} {\eta _{k,t'}^i} }},
\end{align}
to mitigate the error caused by some inaccurate $\eta_{k,t}$ and improve the robustness of the EM algorithm in our proposed OAMP-MMV-ASL algorithm.

\vspace{-2mm}
\subsection{Derivation of Message Update for ASL Model}\label{AP.2}
In the $i$-th iteration of OAMP-MMV-ASL algorithm, we start with the message through the path $x_{k,t} \rightarrow f_{k,t} \rightarrow \alpha_{k}$. Firstly, the message from variable node $x_{k,t}$ to factor node $f_{k,t}$ is ${m^i_{{x_{k,t}} \to {f_{k,t}}}} = {\cal C}{\cal N}\left( {{x_{k,t}};{r^i_{k,t}},{\tau^i_t}} \right)$.
Then, the message from factor node $f_{k,t}$ to variable node $\alpha_{k}$ can be written as
\begin{align}\label{m_ftol}
{m^i_{{f_{k,t}} \to {\alpha_k}}}
& \propto \int {{m^i_{{x_{k,t}} \to {f_{k,t}}}}} {f_{k,t}}\left( {{x_{k,t}},{\alpha_k}} \right)d{x_{k,t}} \notag \\
&= \frac{{1 - {\alpha_k}}}{{\pi {\tau^i_t}}}{\exp({ - \frac{{{{\left| {{r^i_{k,t}}} \right|}^2}}}{{{\tau _t^i}}}})} \notag \\
& \ \ \ + \frac{{{\alpha_k}}}{{\pi {\tau^i_t}L}}\sum\limits_{l = 1}^L {{\exp({ - \frac{{{{\left| {{r^i_{k,t}} - {a_l}} \right|}^2}}}{{{\tau^i_t}}}})}} \notag \\
& \propto \left( {1 - {\eta ^i_{k,t}}} \right)\delta \left( { {\alpha_k}} \right) + {\eta^i_{k,t}}\delta \left( {{\alpha _k}-1} \right),
\end{align}
where $\propto$ denotes equality up to a constant scale factor, and 
\begin{equation}\label{yeta}
{\eta^i_{k,t}} = 1 - \frac{1}{{1 + \frac{1}{L}\sum\limits_{l = 1}^L {q^i_{k,t,l}}}}.
\end{equation}

The backward message is calculated through the path $\hbar_{k} \rightarrow \alpha_{k} \rightarrow f_{k,t} \rightarrow x_{k,t}$.
The message from factor node $\hbar_{k}$ to variable node $\alpha_{k}$ is ${m^i_{{\hbar_k} \to {\alpha_k}}} = \left( {1 - {\lambda^{i-1}_k}} \right)\delta \left( {{\alpha_k}} \right) + {\lambda^{i-1}_k}\delta \left( {{\alpha_k} - 1} \right)$.
Next, the message from variable node $\alpha_{k}$ to factor node $f_{k,t}$ is
\vspace{-1mm}
\begin{align}\label{m_ltof}
{m^i_{{\alpha_k} \to {f_{k,t}}}}
& \propto {m^i_{{\hbar_k} \to {\alpha _k}}}\prod\limits_{t' \in \backslash t} {{m^i_{{f_{k,t'}} \to {\alpha_k}}}} \notag \\
& = {\left(1-\lambda^{i-1}_k\right)}\prod\limits_{t' \in \backslash t} ({{1 - {\eta^i_{k,t'}}})\delta \left( {{\alpha_k}} \right)} \notag \\
& \ \ \ + \lambda^{i-1}_k\prod\limits_{t' \in \backslash t} { {\eta^i_{k,t'}} \delta \left( {{\alpha _k}-1} \right)} \notag \\
& \propto \left(1-{\xi^i_{k,t}}\right)\delta \left( {{\alpha _k}} \right) + {\xi^i_{k,t}}\delta \left( {{\alpha _k}} -1 \right),
\end{align}
where
\begin{equation}\label{kexi}
{\xi^i_{k,t}} = \frac{{\lambda _k^{i-1}\prod\limits_{t' \in \backslash t} {\eta _{k,t'}^i} }}{{\left( {1 - \lambda _k^{i-1}} \right)\prod\limits_{t' \in \backslash t} {\left( {1 - \eta _{k,t'}^i} \right)} + \lambda _k^{i-1}\prod\limits_{t' \in \backslash t} {\eta _{k,t'}^i} }}.
\end{equation}

Finally, the message from factor node $f_{k,t}$ to variable node $x_{k,t}$ is
\begin{align}\label{m_ftox}
\vspace{-1mm}
{m^i_{{f_{k,t}} \to {x_{k,t}}}}
& \propto \sum\limits_{{\alpha_k}} {{m^i_{{\alpha_k} \to {f_{k,t}}}}{f_{k,t}}\left( {{x_{k,t}},{\alpha_k}} \right)} \notag \\
& = \left( {1 - {\xi^i_{k,t}}} \right)\delta \left( {{x_{k,t}}} \right) + \frac{{{\xi^i_{k,t}}}}{L}\sum\limits_{l = 1}^L {\delta \left( {{x_{k,t}} - {a_l}} \right)}.
\end{align}

The approximate posterior distribution of $x_{k,t}$ is given by
\begin{equation}\label{m_post}
p\left( {{x_{k,t}}|{\bf{r}}_k^i} \right) \propto m_{{f_{k,t}} \to {x_{k,t}}}^im_{{x_{k,t}} \to {f_{k,t}}}^i.
\end{equation}



\begin{IEEEbiography}[{\includegraphics[width=1in,height=1.25in,clip,keepaspectratio]{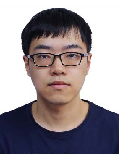}}]{Yikun Mei}
received the B.S. degree from the Beijing Insititue of Technology, Beijing in 2019. He is currently pursuing the Master degree in the Schoold of Inofrmation and Electronics of Beijing Insititue of Technology, Beijing, China. His research interests include massive access for mMTC and sparse signal processing, etc.
\end{IEEEbiography}

\begin{IEEEbiography}[{\includegraphics[width=1in,height=1.25in,clip,keepaspectratio]{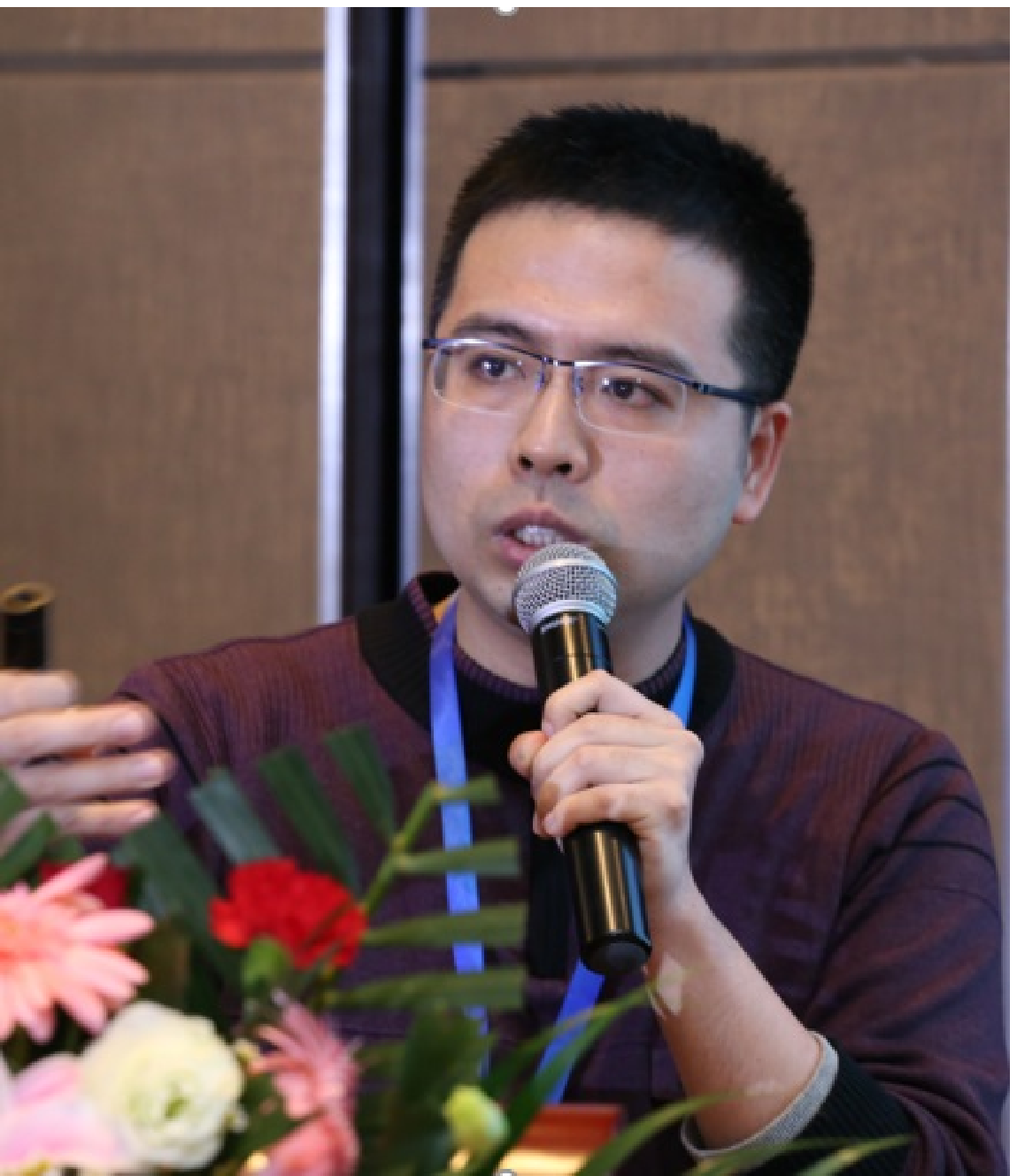}}]{Zhen Gao}
received the B.S. degree in information engineering from the Beijing Institute of Technology, Beijing, China, in 2011, and the Ph.D.
degree in communication and signal processing with the Tsinghua National Laboratory for Information Science and Technology, Department of Electronic
Engineering, Tsinghua University, China, in 2016.
He is currently an Assistant Professor with the Beijing Institute of Technology. His research interests are in wireless communications, with a focus on multi-carrier modulations, multiple antenna systems, and sparse signal processing.
He was a recipient of the IEEE Broadcast Technology Society 2016 Scott Helt Memorial Award (Best Paper), the Exemplary Reviewer of IEEE \textsc{Communication Letters} in 2016, \emph{IET Electronics Letters} Premium Award (Best Paper) 2016, and the Young Elite Scientists Sponsorship Program (2018--2021) from China Association for Science and Technology.
\end{IEEEbiography}

\begin{IEEEbiography}[{\includegraphics[width=1.05in,height=1.3in,clip,keepaspectratio]{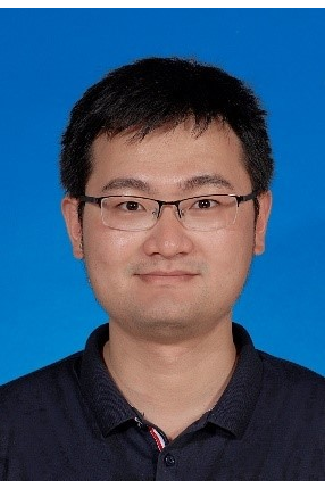}}]{Yongpeng Wu}(S'08--M'13--SM'17)
received the B.S. degree in telecommunication engineering from Wuhan University, Wuhan, China, in July 2007, the Ph.D. degree in communication and signal processing with the National Mobile Communications Research Laboratory, Southeast University, Nanjing, China, in November 2013.

Dr. Wu is currently a Tenure-Track Associate Professor with the Department of Electronic Engineering, Shanghai Jiao Tong University, China. Previously, he was senior research fellow with Institute for Communications Engineering, Technical University of Munich, Germany and the Humboldt research fellow and the senior research fellow with Institute for Digital Communications, University Erlangen-N$\ddot{u}$rnberg, Germany. During his doctoral studies, he conducted cooperative research at the Department of Electrical Engineering, Missouri University of Science and Technology, USA.
His research interests include massive MIMO/MIMO systems, massive machine type communication, physical layer security, and signal processing for wireless communications.

Dr. Wu was awarded the IEEE Student Travel Grants for IEEE International Conference on Communications (ICC) 2010, the Alexander von Humboldt Fellowship in 2014, the Travel Grants for IEEE Communication Theory Workshop 2016, the Excellent Doctoral Thesis Awards of China Communications Society 2016£¬the Exemplary Editor Award of  IEEE Communication Letters 2017, and Young Elite Scientist Sponsorship Program by CAST 2017. He was an Exemplary Reviewer of the IEEE Transactions on Communications in 2015, 2016,2018. He was the lead guest editor for the special issue ``Physical Layer Security for 5G Wireless Networks" of the IEEE Journal on Selected Areas in Communications and lead guest editor of the special issue `` Safeguarding 5G-and-Beyond Networks with Physical Layer Security" of IEEE Wireless Communications. He is currently an editor of the IEEE Wireless Communications, IEEE Transactions on Communications and IEEE Communications Letters. He has been a TPC member of various conferences, including Globecom, ICC, VTC, and PIMRC, etc.
\end{IEEEbiography}

\begin{IEEEbiography}[{\includegraphics[width=1in,height=1.25in,clip,keepaspectratio]{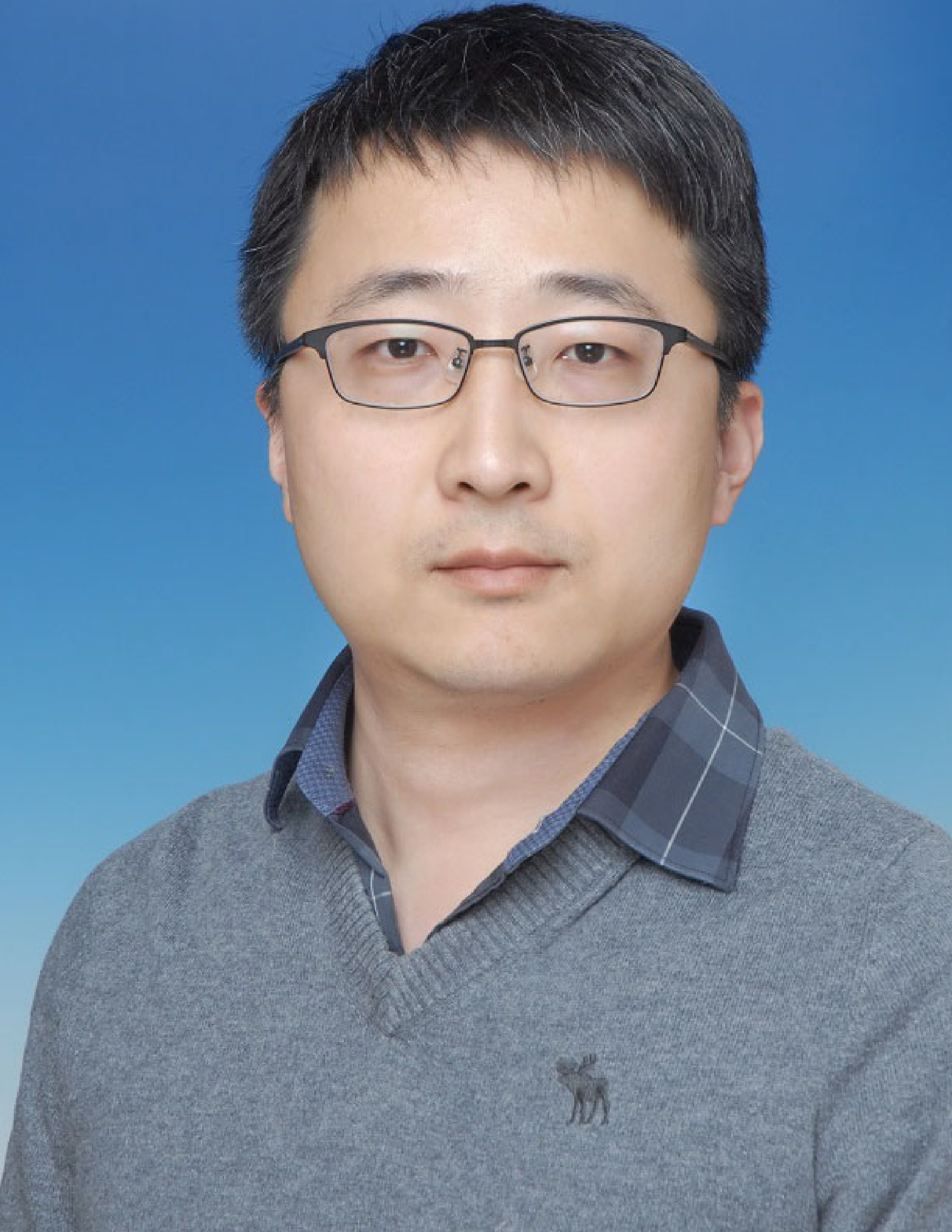}}]{Wei Chen}
(M'13-SM'18) received the B.Eng. degree and M.Eng. degree from Beijing University of Posts and Telecommunications, China, in 2006 and 2009, respectively, and the Ph.D. degree in Computer Science from the University of Cambridge, UK, in 2013. Later, he was a Research Associate with the Computer Laboratory, University of Cambridge from 2013 to 2016. He is currently a Professor with Beijing Jiaotong University, Beijing, China. He is the recipient of the 2013 IET Wireless Sensor Systems Premium Award and the 2017 International Conference on Computer Vision (ICCV) Young Researcher Award. His current research interests include sparse representation, Bayesian inference, wireless communication systems and image processing.
\end{IEEEbiography}

\begin{IEEEbiography}[{\includegraphics[width=1in,height=1.25in,clip,keepaspectratio]{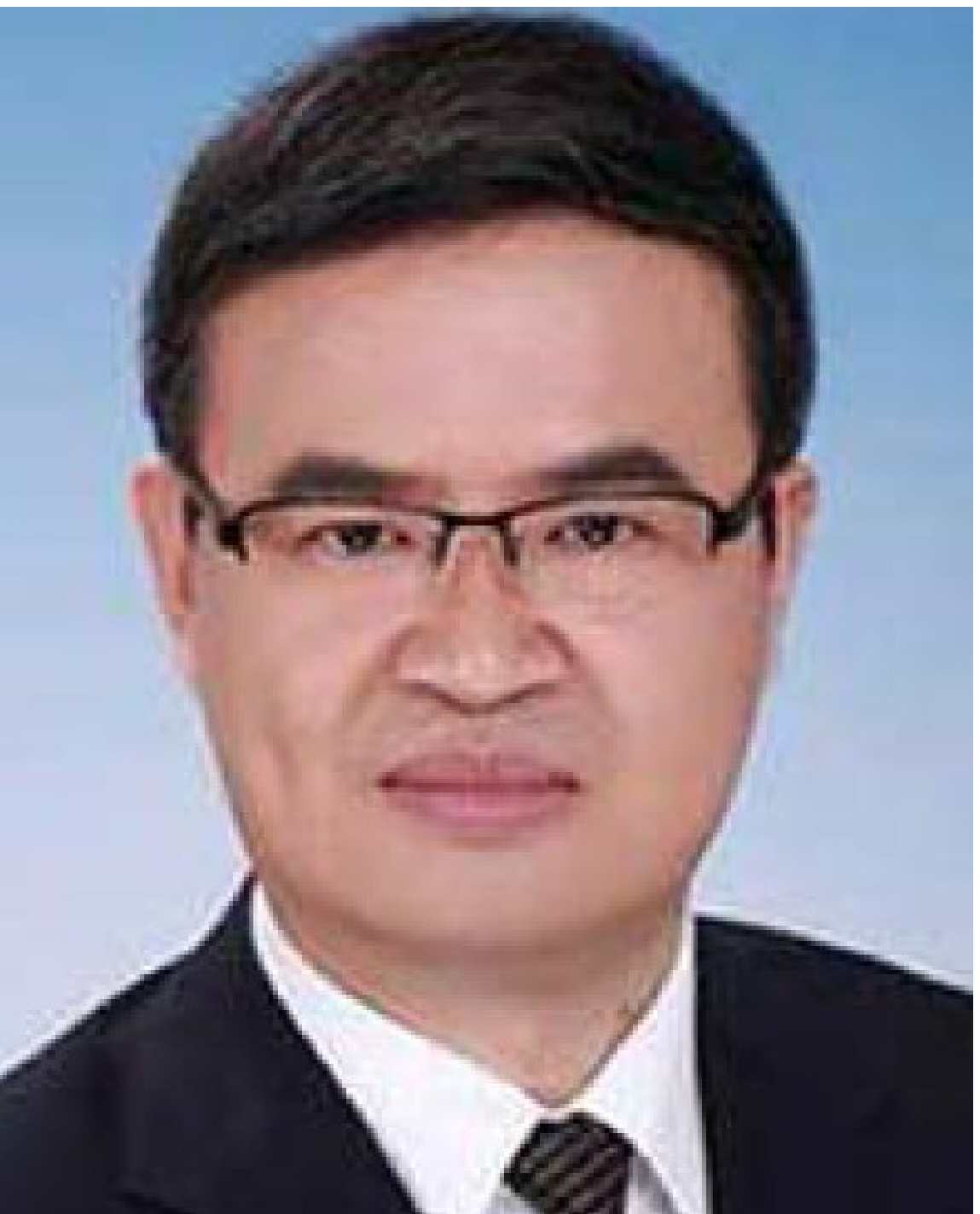}}]{Jun Zhang}
received the B.S., M.S. and Ph.D. degrees in communications and electronic systems from Beihang University, Beijing, China, in 1987, 1991, and 2001, respectively. He used to be a professor in Beihang University, and has served as the dean of the school of Electronic and Information Engineering, the vice president, and the secretary of the Party Committee of Beihang University. Now he is a professor in Beijing Institute of Technology, and also the president of Beijing Institute of Technology.

His research interests are networked and collaborative air traffic management systems, covering signal processing, integrated and heterogeneous networks, and wireless communications. He has won the awards for science and technology in China many times, and he is a member of Chinese Academy of Engineering.
\end{IEEEbiography}

\begin{IEEEbiography}[{\includegraphics[width=0.95in,height=1.4in]{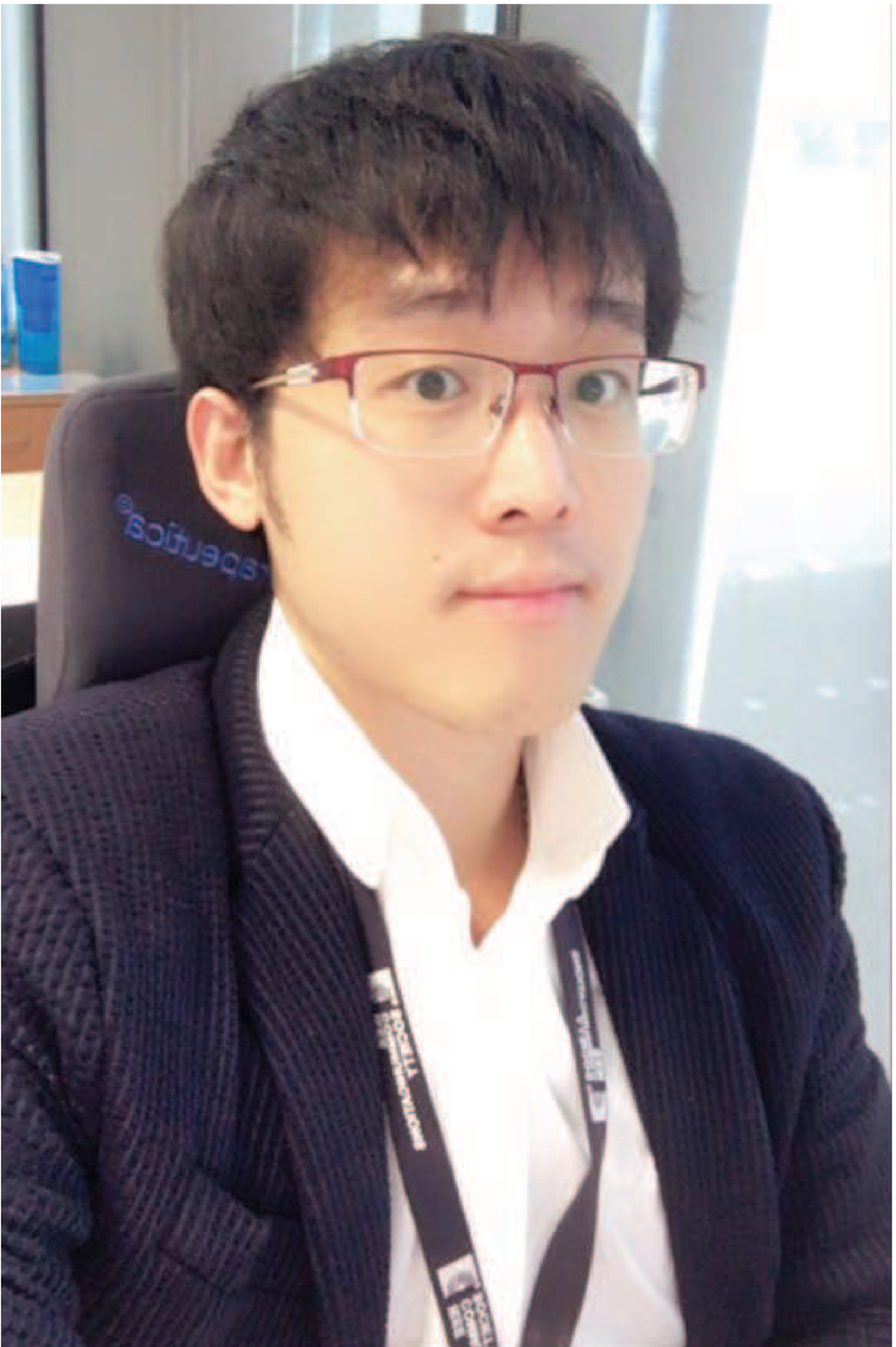}}]{Derrick Wing Kwan Ng}(S'06-M'12-SM'17-F'21) received the bachelor degree with first-class honors and the Master of Philosophy (M.Phil.) degree in electronic engineering from the Hong Kong University of Science and Technology (HKUST) in 2006 and 2008, respectively. He received his Ph.D. degree from the University of British Columbia (UBC) in 2012. He was a senior postdoctoral fellow at the Institute for Digital Communications, Friedrich-Alexander-University Erlangen-N\"urnberg (FAU), Germany. He is now working as a Senior Lecturer and a Scientia Fellow at the University of New South Wales, Sydney, Australia.  His research interests include convex and non-convex optimization, physical layer security, IRS-assisted communication, UAV-assisted communication, wireless information and power transfer, and green (energy-efficient) wireless communications.

Dr. Ng received the Australian Research Council (ARC) Discovery Early Career Researcher Award 2017,   the Best Paper Awards at the WCSP 2020,  IEEE TCGCC Best Journal Paper Award 2018, INISCOM 2018, IEEE International Conference on Communications (ICC) 2018, 2021,  IEEE International Conference on Computing, Networking and Communications (ICNC) 2016,  IEEE Wireless Communications and Networking Conference (WCNC) 2012, the IEEE Global Telecommunication Conference (Globecom) 2011, and the IEEE Third International Conference on Communications and Networking in China 2008.  He has been serving as an editorial assistant to the Editor-in-Chief of the IEEE Transactions on Communications from Jan. 2012 to Dec. 2019. He is now serving as an editor for the IEEE Transactions on Communications,  the IEEE Transactions on Wireless Communications, and an area editor for the IEEE Open Journal of the Communications Society. Also, he has been listed as a Highly Cited Researcher by Clarivate Analytics since 2018.
\end{IEEEbiography}

\begin{IEEEbiography}[{\includegraphics[width=1in,height=1.25in,clip,keepaspectratio]{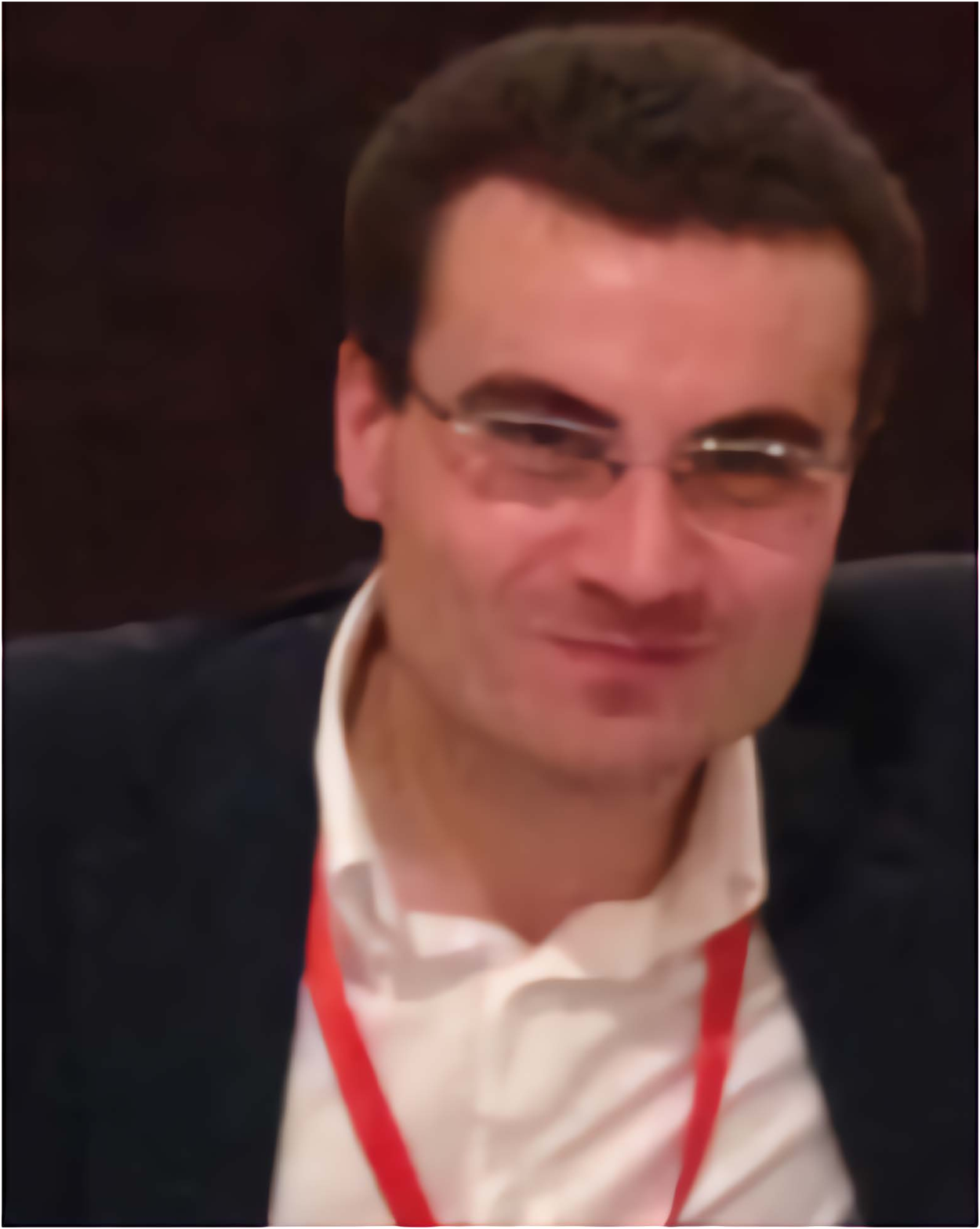}}]{Marco Di Renzo} (Fellow, IEEE)
received the Laurea (cum laude) and Ph.D. degrees in electrical engineering from the University of L'Aquila, Italy, in 2003 and 2007, respectively, and the Habilitation \`{a} Diriger des Recherches (Doctor of Science) degree from University Paris-Sud, France, in 2013. Since 2010, he has been with the French National Center for Scientific Research (CNRS), where he is a CNRS Research Director (CNRS Professor) with the Laboratory of Signals and Systems (L2S) of ParisSaclay University -- CNRS and CentraleSupelec, Paris, France. In Paris-Saclay University, he serves as the Coordinator of the Communications and Networks Research Area of the Laboratory of Excellence DigiCosme, and as a member of the Admission and Evaluation Committee of the Ph.D. School on Information and Communication Technologies. Currently, he serves as the Editor-in-Chief of IEEE Communications Letters, and is a Distinguished Speaker of the IEEE Vehicular Technology Society. In 2017-2020, he was a Distinguished Lecturer of the IEEE Vehicular Technology Society and IEEE Communications Society. Also, he served as an Editor and the Associate Editor-in-Chief of IEEE Communications Letters, and as an Editor of IEEE Transactions on Communications and IEEE Transactions on Wireless Communications. Currently, he serves as the Founding Chair of the Special Interest Group on Reconfigurable Intelligent Surfaces of the Wireless Technical Committee of the IEEE Communications Society, and is the Founding Lead Editor of the IEEE Communications Society Best Readings in Reconfigurable Intelligent Surfaces. He is a Highly Cited Researcher (Clarivate Analytics, 2019), a World's Top 2\% Scientist from Stanford University (2020), a Fellow of the IEEE (2020), a Fellow of the IET (2020), an Ordinary Member of the EASA (2021), and a Member of the Academia Europaea (2021). He has received several individual distinctions and research awards, which include the IEEE Communications Society Best Young Researcher Award for Europe, Middle East and Africa, the Royal Academy of Engineering Distinguished Visiting Fellowship, the IEEE Jack Neubauer Memorial Best System Paper Award, the IEEE Communications Society Young Professional in Academia Award, the SEE-IEEE Alain Glavieux Award, the 2019 IEEE ICC Best Paper Award, the Nokia Foundation Visiting Professorship, and the 2021 EURASIP Journal on Wireless Communications and Networking Best Paper Award.
\end{IEEEbiography}
\end{document}